\documentclass[useAMS,usenatbib]{mn2e}

\usepackage{natbib}
\usepackage{url}
\usepackage{setspace}
\usepackage{fancyhdr}
\usepackage{graphicx}
\usepackage{makeidx}
\usepackage{amsmath}
\usepackage{subfigure}

\usepackage{txfonts} 
\usepackage{lscape} 
\usepackage{longtable}
\usepackage{rotating}

\usepackage{booktabs}
\usepackage{tabularx}

\newcommand{\apjs}{ApJS}
\newcommand{\apjl}{ApJL}
\newcommand{\aap}{A{\&}A}

\newcommand{\chs}{{$\rm{CH{_4}s}$}}
\newcommand{\chl}{{$\rm{CH{_4}l}$}}

\title[Methane imaging]{49 new T~dwarfs identified using methane imaging}
\author[C. V. Cardoso et al.]{C. V. Cardoso$^{1, 2}$, B. Burningham$^{2,3}$\thanks{E-mail:
B.Burningham@herts.ac.uk}, R. L. Smart$^{1}$, L. van Spaandonk$^{2,4}$, D. Baker$^{2}$, L. C. Smith$^{2}$, 
\newauthor 
A. H. Andrei$^{5,2,6,7}$, B. Bucciarelli$^{1}$, S. Dhital$^{8}$, H. R. A. Jones$^{2}$, M. G. Lattanzi$^{1}$, \\
\newauthor
 A. Magazz\'u$^{9}$, D. J. Pinfield$^{2}$, C. G. Tinney$^{10,11}$\\
$^{1}$ Istituto Nazionale di Astrofisica, Osservatorio Astrofisico di Torino, Strada Osservatorio 20, 10025 Pino Torinese, Italy \\
$^{2}$ Centre for Astrophysics Research, Science and Technology Research Institute, University of Hertfordshire, Hatfield AL10 9AB, UK\\
$^{3}$  NASA Ames Research Center, Mail Stop 245-3, Moffett Field, CA 94035, USA \\
$^{4}$ King Edward VI School, Church Street, Stratford-upon-Avon, Warwickshire, CV37 6HB, UK\\
$^{5}$ Observat\'orio Nacional,  Rua General Jos\'e Cristino, 77 - S\~ao Crist\'ov\~ao, Rio de Janeiro - RJ, 20921-400, Brazil\\
$^{6}$ SYRTE, Observatoir de Paris, 61 Avenue de l'Observatoire, 75014 Paris, France
$^{7}$ Observat\'orio do Valongo/UFRJ, Ladeira Pedro Antonio 43, Rio de Janeiro - RJ, 20080-090, Brazil\\
$^{8}$ Department of Astronomy, Boston University,725 Commonwealth Ave, Boston MA 02215, US\\
$^{9}$ Fundaci\'on Galileo Galilei-INAF, Rambla J. A. Fern\'andez  P\'erez 7, 38712 Bre\~na Baja, Spain\\
$^{10}$  Australian Centre for Astrobiology, University of New South Wales, 2052, Australia \\
$^{11}$  School of Physics, University of New South Wales, 2052, Australia}

\begin{document}

\maketitle

\label{firstpage}

\begin{abstract}

We present the discovery of 49 new photometrically classified T dwarfs from the combination of large infrared and
optical surveys combined with follow-up TNG photometry. We used multi-band infrared
and optical photometry from the UKIRT and Sloan Digital Sky Surveys to identify
possible brown dwarf candidates, which were then confirmed using methane filter photometry.
We have defined a new photometric conversion between CH$_4$s - CH$_4$l colour and spectral type for T4 to T8 brown dwarfs based on a part of the sample that has been followed up using methane photometry and spectroscopy. Using methane differential photometry as a proxy for spectral type for T~dwarfs has proved to be a very efficient technique. Of a subset of 45 methane selected brown dwarfs that were observed spectroscopically, 100\% were confirmed as T~dwarfs.
Future deep imaging surveys will produce large samples of faint brown dwarf candidates, for which spectroscopy will not be feasible. When broad wavelength coverage is unavailable, methane imaging offers a means to efficiently classify candidates from such surveys using just a pair of near-infrared images.

\end{abstract}

\begin{keywords}
surveys -- brown dwarfs -- stars: low-mass
\end{keywords}

\section[]{Introduction}

Brown dwarfs are intrinsically faint objects that cool with time, because they are unable to sustain a permanent nuclear energy source. Their temperatures range between a few hundred Kelvin and a few thousand Kelvin, with T~dwarfs being cooler than $\sim1500$K, with most of their energy emitted at the infrared wavelengths. 
\cite{Tsuji1964} predicted that at these temperatures brown dwarfs would present strong methane absorption in their atmospheres. The discovery of the first 
T dwarf \citep[Gl 229B; ][]{nakajima1995} confirmed this prediction. 
The spectral classification of T~dwarfs uses the presence and strength of methane absorption lines in their atmospheres as a key feature for their classification \citep{burgasser06}.

Due to their inherent faintness, spectroscopic classification of a statistically useful sample (e.g. 100s) of T~dwarfs across the full sequence requires large amounts of time, typically on 8m-class telescopes. While the new generation of wide-field surveys will deliver the ability to study the Galactic substellar population with previously unachievable statistical power, the extreme faintness and large number of the candidates produced make spectroscopic follow-up of such samples infeasible.  
For example, the Dark Energy Survey \citep[DES; ][]{des} will cover some 5000 square degrees of sky to a full depth of $z'_{AB} \approx 23.5$, representing a factor $>30$ increase in volume over that searched by the Sloan Digital Sky Survey \citep[SDSS; ][]{sdss}, which alone discovered around 50 T~dwarfs, and played a complementary role is discovering many more. Further down the line, the Large Synoptic Survey Telescope \citep[LSST][]{lsst} and Euclid  \citep{euclid} projects will combine to provide a factor 4000--8000 increase in the searchable volume for T~dwarfs compared to the hugely successful 2~Micron All Sky Survey \citep[2MASS][]{2mass}, which discovered $\sim$100 T~dwarfs. 

The vast majority of these 10,000s of brown dwarfs will be fainter than the current spectroscopic reach of 8m telescopes, so developing reliable and accurate photometric classification methods is highly desirable, and notable progress has been made at classifying ultracool dwarfs using multiwavelength photometry by \citet{skrzypek2014}. Effective as this method is, it relies on well measured photometry over a wide wavelength range, and has thus been largely restricted to the brighter end of the brown dwarf population where there is good overlap between optical, near-infrared and longer wavelength surveys. However, much of the extra volume explored in recent large area surveys such as the UKIRT Infrared Deep Sky Survey \citep[UKIDSS; ][]{ukidss} and VISTA Hemispheres Survey (VHS; PI McMahon, Cambridge, UK) has survey wavelength coverage limited to as few as two near-infrared bands. Selection of cool brown dwarfs in this domain is typically facilitated by constraints on photometry provided by ``drop-out" methods \citep[see e.g. ][]{ben2013}, and any further photometric classification must thus rely on follow-up imaging.

Classification of T~dwarfs using filters centered on the methane absorption bands provides the opportunity to achieve this with just one or two additional images for each source (depending on whether $H$ band photometry is already available) .
Methane imaging has already proved a valuable technique for detecting T~dwarfs. It has been used in the search for T~dwarfs in star forming regions \citep[e.g., ][]{2009A&A...508..823B, parker2013}, in the field \citep[e.g., ][]{tinney2005} and in search of planetary companions \citep[e.g., ][]{2013ApJ...777..160B}. 
  In this work we describe our use of methane filters to perform a spectral classification of T~dwarfs selected from the UKIDSS Large Area Survey (LAS). 
  This classification formed a key part of the candidate selection and prioritisation 
process employed to discover 76 new T dwarfs in the UKIDSS LAS by  \citet[][ hereafter BCS13]{ben2013}.
In addition to selecting candidates that were confirmed by spectroscopy and published in BCS13, this process identified a further 49 targets with methane colours consistent with T~dwarfs status, which we present here for the first time.

\begin{figure}
  \resizebox{\hsize}{!}{\includegraphics{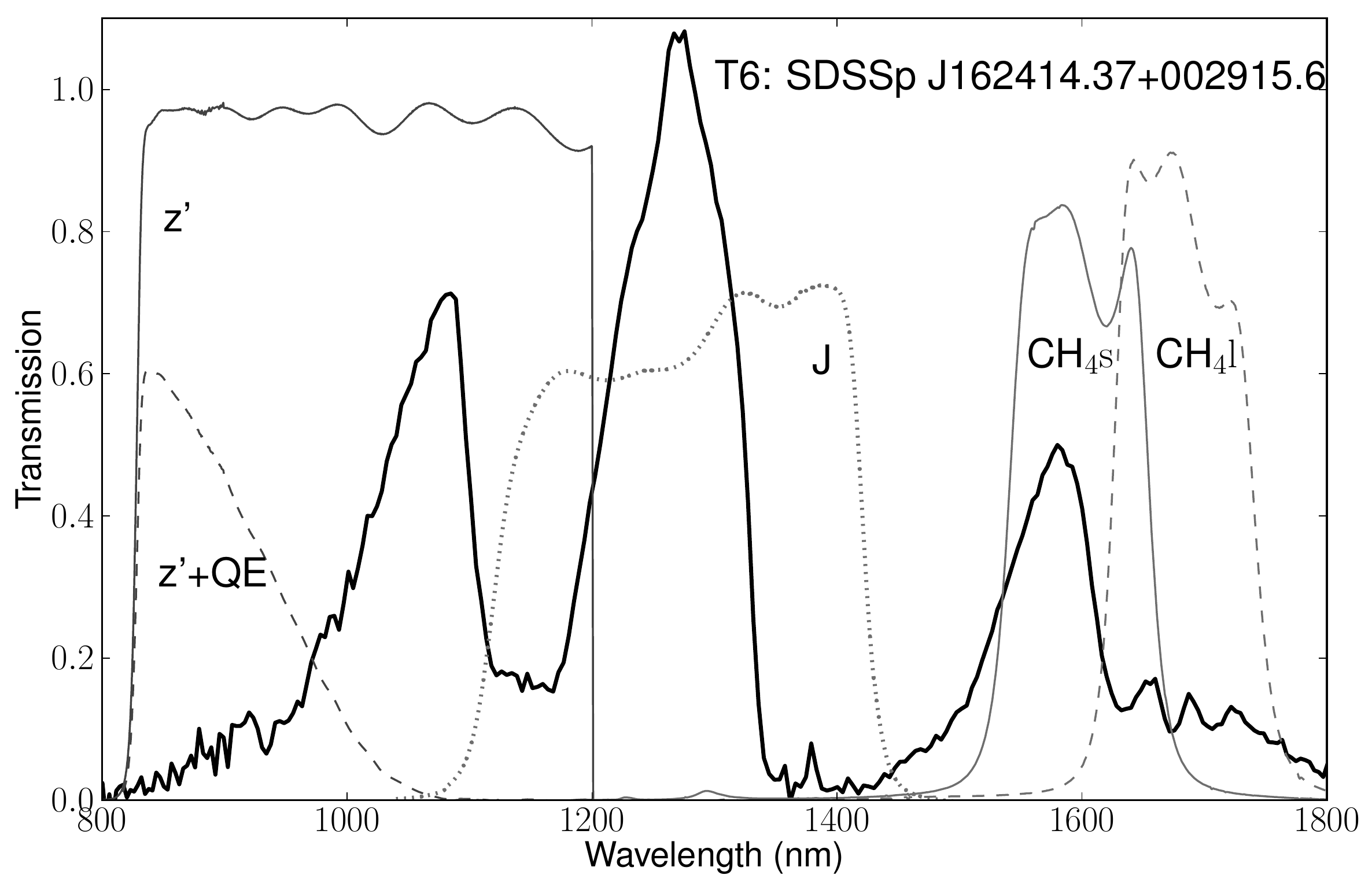}}
 \caption[TNG filters and T6 BD spectrum.]{The filter profiles of the filters used in our follow-up of T~dwarf candidates. The z$^\prime$ filter is shown as the filter response (full dark grey line), and the filter response when convolved with the CCD quantum efficiency (QE, dashed dark grey line) as the latter defines the red cut-off.  The image also shows the spectrum of T6 SDSSp~J162414.37+002915.6 (thick black line, data retrieved from SpeX Prism Spectral Libraries\footnotemark, originally published by \citealt{Burgasser:2006p7181}, the spectrum is normalized at 1.27\micron).}
\label{fig:filters}
\end{figure}

Our strategy exploits the prominent methane absorption feature present at $\sim1.58\micron$ by using two adjacent filters: one that lies on this feature (CH$_4$s); and another that lies close to the continuum (CH$_4$l). In Figure~\ref{fig:filters} we show the profiles of the near-infrared filters used along with the spectrum of a T6 dwarf. These filters match those used by \citet{tinney2005}, and we base our work on the same methane photometric system.

The combination of methane imaging with preselection of candidates by broadband photometry provides a robust method of identifying and classifying T~ dwarfs.  Of the targets photometrically classified as T dwarfs by the method outlined in this work that were followed up spectroscopically in BCS13, 100\% were confirmed as T~dwarfs with spectral types within the range expected by the uncertainties.  Thus, although the targets discovered in this work are photometrically classified as T~ dwarfs, this photometric classification can reasonably be seen as equivalent to more traditional spectral typing, and the language we use for the remainder of this paper reflects this view.

In Section~\ref{sec:candsel} we briefly summarise the candidate selection method. In Section~\ref{sec:followup} we describe our photometric follow-up observations, the calibration of our methane photometry. In Section~\ref{sec:class} we discuss the classification of our targets, and highlight new discoveries. In Section~\ref{sec:binary} we discuss our search for wide common proper motion binaries amongst the newly discovered T~dwarfs. We summarise our results in Section~\ref{sec:summary}.

\section[]{Candidates selection procedure}
\label{sec:candsel}

The photometry described in this paper was carried out as part of the candidate selection and prioritisation methodology  previously presented in BCS13. \footnotetext{\url{http://pono.ucsd.edu/~adam/browndwarfs/spexprism/}}
Here we present a brief summary of the selection procedure, and we direct the reader to BCS13 and references therein for a more complete description and justification of the process. 
Our initial selection of T~dwarf candidates is based on two sets of colour cuts using the data from UKIDSS, which covers the $YJHK$ bands of the Mauna Kea Observatories system \citep[MKO; ][]{mko},  and the SDSS, which covers the $u'g'r'i'z'$ bands. 
We initially select objects which have $J-H < 0.1$, and either no detection in the K-band, or $J-K < 0.1$.
We then cross-match this selection against SDSS, and only accept as candidates those objects which are either undetected in SDSS within 2~arcseconds of the UKIDSS source, or have $z^\prime - J > 2.5$. 
Since the Large Area Survey (LAS) has nominal 5$\sigma$ limits of 18.8 and 18.2 in the H and K bands respectively (c.f 20.2 and 19.6 for Y and J), we ensure that we are complete to our faint cut-off at $J=18.8$ by further selecting sources that are detected in Y and J, but are undetected in $H$ and $K$, with $Y - J > 0.5$ for $J < 18.5$. 

Contamination of our sample is dominated by M~dwarfs and by solar system objects (SSOs), which can display similar colours to very cool T~dwarfs in the near-infrared, whilst SSOs also appear as drop-outs in SDSS due to their fast motion.
The YJ only selection is particularly vulnerable to contamination by fast-moving SSOs since the LAS Y and J observations were typically carried out as pair in the same observing block, but not necessarily on the same night
as the H and K images (also acquired in a single observing block).
As a result, a moving SSO might appear in the $Y$ and $J$ images taken contemporaneously, but not in the $HK$ images taken at a different time. 
Extragalactic objects do not represent a significant source of contamination in our search. The only extragalactic sources likely to lie within our $z'YJHK$ selection are extremely high-redshift ($z > 6$) quasars \citep[e.g. ][]{hewett06}, which have a very low sky-surface density \citep[see e.g. ][]{mortlock2012}. So, whilst much effort is required to remove substellar contaminants from extragalactic searches in this colour space, the reverse is not true.

We have not included a photometric cut based on WISE photometry because, as shown by BCS13 (Figures 5 and 6), the faint end of UKIDSS probes a much greater depth for early-~to mid-T~dwarfs than WISE. For example, the maximum depth of ALLWISE is $W1 \approx 17.8$ and $W2 \approx 16.5$. For a T5 dwarf $J - W1 \approx 0$, and $J - W2 \approx 1-2$. So, although a WISE detection for a source in the $J > 18.0$ regime might suggest a later spectral type, a non-detection does not help discriminate a bona-fide mid-T candidate from an M~dwarf contaminant, which would also be undetected.

\section[]{Photometric follow-up}
\label{sec:followup}

The objects that survive this initial selection were then followed up as part of a large program to detect and characterise late T~dwarfs at the 3.5m Telescopio Nazionale Galileo (TNG) on Roque de Los Muchachos Observatory (ORM, La Palma, Spain).
The observations were taken between 2010 and 2013 under the INAF TNG Very Large Program proposals:
AOT22 TAC 96, AOT23 TAC 28, AOT24 TAC 49, AOT25 TAC 32 and AOT26 TAC
68. 

Using the candidate list defined in section 2.1 we adopted a 3 step photometric follow-up procedure at the TNG:

{\bf (1) Screening out SSOs and faint J band scatters:} Candidates are initially re-observed in the J-band to either yield a SNR=10-20 detection, or reject the candidate as an SSOs (which is not detected at this
second epoch). Roughly $30\%$ of our candidates are ruled out at this stage, split roughly evenly between the YJH and YJ only samples.

{\bf (2) Removal of M dwarf contaminants:} 
Candidates that pass step (1) and were not detected in SDSS were then observed in the $\rm{z}^\prime$-band.
This step allows us to reject M dwarfs that will have colour of $\rm{z}^\prime-J < 2.0$, at our $J=18.8$ cut off limit. 
Approximately $30\%$ of the remaining candidates with $J < 18.8$ were ruled out at this stage, with the proportion of contaminants rising steeply at fainter magnitudes.

{\bf (3) Differentiating between early and late-T~dwarfs:} 
The remaining candidates (which should be primarily composed of late L and T~dwarfs) were then imaged using the CH$_4$s and CH$_4$l  filters. 
The resulting methane colour allows us estimate spectral types for the T dwarfs following the method of \cite{tinney2005}. 
Of the 116 targets that we observed with methane imaging, 22 were ruled out as non- T~dwarfs by their methane colours (see Section~\ref{sec:class}).

A tabular summary of the observations is given in Appendix~Table~\ref{tab:observations}.

\subsection[]{J band Imaging}

Infrared photometry was obtained using NICS, the Near Infrared Camera Spectrometer \citep{2001A&A...378..722B} mounted at the Cassegrain focus of TNG.
The data used in this work was acquired with the large field mode that has a field of view of $4.2 \times 4.2$ arcmin and a pixel scale of 0.25 arcsec/pixel.

To exclude SSO interlopers within the T~dwarf selection, short (300s integration) $J$ band observations were obtained to confirm the presence and location of targets with single epoch detections in UKIDSS. To allow effective background subtraction, a 5-point dither pattern with 2$\times$30 second integrations at each location was used.
We reduced and combined the images into mosaics using the reduction Speedy Near-infrared data Automatic Pipeline (SNAP version 1.3\footnote{\url{http://www.tng.iac.es/news/2002/09/10/snap/}}) provided by the TNG. SNAP is an automated wrapper of existing pieces of software (IRDR, IRAF, Sextractor and Drizzle) to perform a full reduction with a single command.
SNAP performs flat-fielding, computes the offsets between the dithered images, creates a mosaic image with double-pass sky subtraction and correction for field distortion.

\subsection[]{Optical Photometry with LRS at TNG}

$\rm{z}^\prime$ photometry was taken using DOLORES (the Device Optimized for LOw RESolution, hereafter LRS in short) at TNG \citep{1997MmSAI..68..231M}.
LRS is equipped with a  $2048 \times 2048$ pixels CCD with a field-of-view of $8.6 \times 8.6$ arcmin with a 0.252 arcsec/pixel scale. Figure~\ref{fig:filters} shows the $\rm{z}^\prime$ Sloan filter's profile and the resultant profile of the convolution of the filter with the CCD's quantum efficiency. Based on the objectÕs magnitude and the weather conditions
the observations were made with a long exposure and two shorter undithered exposures.
For each night a set of standard calibration flatfields and dark observations were taken. The images were dark subtracted and flatfielded, and (where appropriate) combined using standard IRAF routines. No attempt was made to defringe these images -- the E2V4240 CCD detector in LRS has a low fringing level, and the science object was usually positioned in the top right section of the CCD where fringing is minimal.
The data were taken in a variety of observing conditions, from photometric conditions to cirrus. The photometric zero point was calibrated using the non-saturated SDSS stars present in the field of view. Photometry was performed with IRAF using a fixed circular aperture of radius 2$\arcsec$.  As described in BCS13, the response of the DOLORES detector coupled with its SDSS $z'$ filter is sufficiently similar to the SDSS equipment that no correction was applied to the SDSS database magnitudes.

The purpose of the $z'$ band imaging was principally to establish the red $z' - J$ colour of candidate T~dwarfs, and screen out bluer (and more common) M~dwarfs. It was only obtained for fainter targets that lacked a useful $z' - J$ constraint from SDSS and/or lacked a UKIDSS $H$ band detection. Integration times were set so that a non-detection would indicate a T~dwarf-like $z' - J > 2.5$. Good signal-to-noise $z'$ band imaging was not the aim, and where detections were made the resulting magnitudes have large uncertainties.

\subsection[]{Methane Photometry}

\subsubsection{Observations}

NICS is equipped with a set of Mauna Kea Observatories near-infrared CH$_4$s and CH$_4$l filters as specified by \citet{2002PASP..114..180T} (as shown in Figure~\ref{fig:filters}), and 1024x1024 Rockwell HAWAII-1 HgCdTe detectors. 

Photometry in these two methane filters provides information about the strength of the methane absorption bands in T~dwarfs. CH$_4$s samples the methane absorption bands present between 1.5 and 1.7 \micron, while the CH$_4$l samples a pseudo-continuum outside the methane band. 
The methane absorption bands become more prominent for later T dwarfs and this can be quantified using the difference between observed CH$_4$s and {CH$_4$l magnitudes. 
Most of the data were taken with a 30 pointing dither pattern and a total integration time varying between 20 minutes and 1 hour, with the individual integrations not larger than 30 seconds, under conditions varying from photometric to cirrus cloud cover.

\subsubsection[]{Zeropoint calibration}
\label{sec:ch4cal}

The methane photometry was not taken in photometric conditions, and during this program no photometric standards were observed.
Photometric calibration was achieved using UKIDSS and SDSS photometric information for 
background stars in the NICS field-of-view, following the procedure outlined in \cite{tinney2005}.  

\citet{tinney2005}  used IRIS2 on the 3.9 m Anglo-Australian Telescope (AAT), which employs a similar set of methane filters ({\chs} and {\chl}) to NICS at TNG and a similar set of infrared 1024x1024 Rockwell HAWAII-1 HgCdTe detectors.  
The \citet{tinney2005} methane system is defined, in absolute terms, by the $H$ band magnitudes of UKIRT faint standard A,F and G type stars for which $H =$~ {\chs}~=~{\chl}.   By observing a wide range of stars and brown dwarfs, along with UKIRT standards, they then derived a transformation between absolute methane colour and spectral type (Equation~2 of \citealt{tinney2005}).

In principal, AFG stars in the field-of-view could be used to trivially calibrate the zero-point in any methane observations if a sufficient number of such stars were available in a given field of view.  However, the $8 \times 8$ arcmin field-of-view of IRIS2 (as for the $8.6 \times 8.6$ arcmin field of NICS) would rarely yield a sufficient batch of AFG stars for this purpose. 
To allow a wider range of spectral types to be employed for differential methane zero-point determination, and increase the number of differential calibrators in any field-of-view,  \citet{tinney2005} also derived a differential (\chs -- \chl)  colour based on 2MASS photometry for background stars, using a calibration from $(J-H)_{2MASS}$ to (\chs -- \chl) (Tinney et al.'s Equations 3 and 4) to correct for colour terms in those background stars.
Our methane photometric calibration follows this outline, but is based on $(J-H)_{MKO}$, rather than $(J-H)_{2MASS}$, since we have UKIDSS MKO photometry for all our fields.

We start our photometric extraction by performing aperture photometry to all the objects in the field using IMCORE (part of CASUTOOLS)  using a fixed circular aperture with $1\arcsec$ radius.
We then cross-identify the methane detections with the UKIDSS LAS survey. The calibration stars needed to have $J$ and $H$ band detections and magnitude errors smaller than 0.1 mag.

\begin{figure}
  \resizebox{\hsize}{!}{\includegraphics{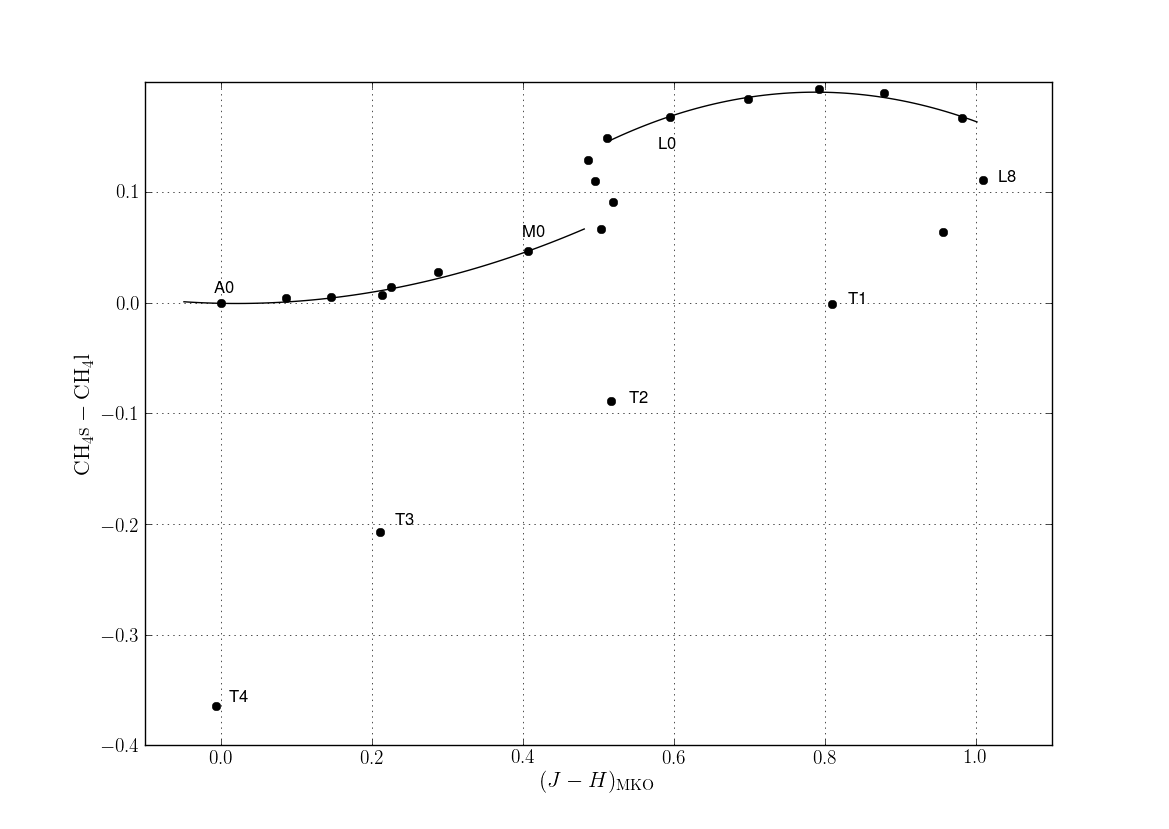}}
 \caption[.]{{\chs} - {\chl} dwarf sequence as a function of $\rm{J-H}$ in the MKO photometric system. The parameterisation curves are presented in Equation\,\ref{eqn:Tinney3} and \ref{eqn:Tinney4} and represent polynomial fits in the ranges $-0.05<\rm{J-H}<0.48$ and $0.51<\rm{J-H}<1.00$ to the data given in Table 3 of \citet{tinney2005}. This parametrization is used in the calibration of the $\rm{CH_4}$ photometric zero point.} 
\label{fig:parametrization_Tinney}
\end{figure}

A first zero point estimate is determined converting the $(J-H)_{\rm{MKO}}$ colour of all objects to a methane colour. 
Equations~3 and~4 of \cite{tinney2005}  give this colour parametrisation for the 2MASS photometric system, we base our photometry on the UKIDSS photometry that is on the MKO photometric system. We have recalculated this parametrisation based on the colour information in the MKO system given in Table 3 of \citet{tinney2005}. Figure~\ref{fig:parametrization_Tinney} shows the parametrisation curve used to convert from $(J-H)_{MKO}$ to calibrated methane colour.
We avoided the region $0.48 < (J-H)_{\rm{MKO}} < 0.512$ where the sequence is degenerate. The sequence was fitted with two separate quadratics to the regions $-0.05<(J-H)_{\rm{MKO}}<0.48$,

\begin{equation}
\label{eqn:Tinney3}
\rm{CH{_4}s} - \rm{CH{_4}l} = +0.00046 - 0.01259 (J-H) +0.31817 (J-H)^2
\end{equation}

and $0.51 <  (J-H)_{\rm{MKO}} < 1.00$, 

\begin{equation}
\label{eqn:Tinney4}
\rm{CH{_4}s} - \rm{CH{_4}l} = -0.17317 + 0.92744 (J-H) -0.58969 (J-H)^2 .
\end{equation}

A first zero point is calculated between instrumental methane colour and calibrated methane colour using a linear fit with a 1-to-1 plus a zero point relationship.
After the first zero point estimation, a cut to exclude objects with methane colours consistent with T~dwarfs is applied, removing objects with {\chs} -- {\chl} $< -0.2$. The linear fit is repeated.
A third cut is then applied to remove objects that differ from the zero point fit more than $2 \sigma$, and the final zero point determined repeating the linear fit.
We can then transform the instrumental methane colour to a calibrated methane colour.

\section[]{Methane classification of T~dwarfs}
\label{sec:class}

\subsection{Calibration of spectral type- {\chl} -- {\chs} relation}.

To obtain an initial estimate of the spectral type we reverse Equation~2 of \cite{tinney2005}, from now on referred to as Equation~\ref{eqn:Tinney2}:

\begin{equation}
\label{eqn:Tinney2}
\begin{split}
\rm{CH{_4}s} - \rm{CH{_4}l} = & n (0.0087 + 6.176\times10^{-6}n^2 \\
                                               & + 1.202\times10^{-9}n^4 - \dfrac{0.519}{68.5 - n} )
\end{split}
\end{equation}

This function is numerically valid to $n < 68.5$, but data were only available on the range $0 < n <62$ that represents the spectral range from A to T (A+0, F+10, G+20, K+29, M+35, L+45, T+54).  We note that there is some overlap of spectral types in some ranges of the index around K,M and late-L spectral types due to historical gaps in the sequence. However since the expression is essentially degenerate at earlier types, and is only used to estimate spectral types for for T0 and later, this issue has no impact on our science.

A large portion of our initial sample of late T~dwarfs were followed up spectroscopically within our program by BCS13, and elsewhere. 
We have obtained methane colors for a total of 40 T~dwarfs that have also been spectroscopically observed, and we present their properties in Table~\ref{tab:spec}  \citep[BCS13, ][]{ben2011b,ben10b,albert2011,kirkpatrick2011,pinfield08,burgasser06}. A further five objects were undetected in {\chl} and so shortlisted as late-T dwarfs, and subsequently confirmed as such by spectroscopy. 
In Figure~\ref{fig:TinneyvsSpec} we compare the classification obtained using methane colours with the published spectroscopic classification. 

\begin{landscape}
\begin{table}
{\scriptsize
\begin{tabular}{ c  >{$}c<{$}  >{$}c<{$}  >{$}c<{$}  >{$}c<{$}  >{$}c<{$}  >{$}c<{$}   c  c  c  c  >{$}c<{$}  >{$}c<{$}  >{$}c<{$}  c}
\hline
Name  & z' & \rm{Y_{MKO}} & \rm{J_{MKO}} & \rm{H_{MKO}} & \rm{K_{MKO}} &  {\rm CH_4s} - {\rm CH_4l} & \multicolumn{3}{c}{${\rm CH_4}$ Photometric Type} & SpType &  {\rm CH_4-J} &  {\rm CH_4-H} &  {\rm CH_4-K} & SpT Reference\\
& & & & & & & Best & Min & Max &  & & & & \\
\hline
ULASJ000734.90+011247.1 & - & 19.22 \pm 0.07 & 18.05 \pm 0.04 & - & - & -0.91 \pm 0.13 & T6.7 & T6.3 & T7.0 & T7 & 0.286\pm 0.07 {\rm \,(T6/7)} & 0.233\pm 0.007 {\rm \,(T7)} & 0.118\pm 0.010 {\rm \,(>T6)} & BCS13\\
ULASJ012735.66+153905.9 & - & 19.47 \pm 0.13 & 18.22 \pm 0.07 & 18.62 \pm 0.14 & - & -0.88 \pm 0.17 & T6.6 & T6.0 & T7.0 & T6.5& 0.307 \pm 0.07 {\rm \,(T6)} & 0.280 \pm 0.011 {\rm \,(T6)} & 0.127 \pm 0.012 {\rm \,(T6/7)} & BCS13\\
ULASJ012855.07+063357.0 & 21.56 \pm 0.16 & 19.66 \pm 0.14 & 18.93 \pm 0.12 & - & - & -0.81 \pm 0.14 & T6.3 & T5.9 & T6.7 & T6& 0.320 \pm 0.008 {\rm \,(T6)} & 0.392 \pm 0.014 {\rm \,(T5)} & 0.172 \pm 0.023 {\rm \,(T5/6)} &  BCS13\\
ULASJ013017.79+080453.9 & 21.74 \pm 0.21 & 19.06 \pm 0.03 & 17.93 \pm 0.02 & 18.21 \pm 0.02 & 18.35 \pm 0.04 & -0.72 \pm 0.07 & T6.0 & T5.8 & T6.2 & T6& 0.338 \pm 0.005 {\rm \,(T6)} & 0.291 \pm 0.009 {\rm \,(T6)} & 0.100 \pm 0.012 {\rm \,(>T6)} & BCS13\\
ULASJ013950.51+150307.6 & - & 19.72 \pm 0.17 & 18.44 \pm 0.10 & 18.53 \pm 0.18 & - & -0.83 \pm 0.11 & T6.4 & T6.1 & T6.7 & T7& 0.286 \pm 0.014 {\rm \,(T6/7)} & 0.209 \pm 0.010 {\rm \,(T7)} & 0.084 \pm 0.019 {\rm \,(>T6)} &BCS13\\
ULASJ020013.18+090835.2 & - & 18.98 \pm 0.07 & 17.81 \pm 0.04 & 18.18 \pm 0.11 & 18.18 \pm 0.20 & -0.83 \pm 0.11 & T6.4 & T6.0 & T6.7 & T6& 0.285 \pm 0.008 {\rm \,(T6/7)} & 0.237 \pm 0.008 {\rm \,(T7)} & 0.079 \pm 0.013 {\rm \,(>T6)} &BCS13\\
ULASJ074502.79+233240.3 & 22.05 \pm 100.00 & 20.00 \pm 0.15 & 18.88 \pm 0.07 & - & - & -1.62 \pm 0.17 & T8.3 & T8.0 & T8.5 & T8.5&0.130 \pm 0.039 {\rm \,(>T7)} & 0.069 \pm 0.012 {\rm \,(>T7)} & 0.102 \pm 0.023 {\rm \,(>T6)} & BCS13\\
ULASJ075937.75+185555.0 & 22.80 \pm - & 20.21 \pm 0.18 & 18.70 \pm 0.07 & - & - & -0.95 \pm 0.12 & T6.8 & T6.5 & T7.1 & T6& 0.324 \pm 0.004 {\rm \,(T6)} & 0.274 \pm 0.008 {\rm \,(T6)} & 0.090 \pm 0.008 {\rm \,(>T6)} &BCS13\\
ULASJ081110.86+252931.8 & - & 18.76 \pm 0.03 & 17.57 \pm 0.02 & 18.19 \pm 0.12 & 18.02 \pm 0.19 & -1.03 \pm 0.13 & T7.0 & T6.7 & T7.3 & T7& 0.259 \pm 0.002 {\rm \,(T7)} & 0.229 \pm 0.003 {\rm \,(T7)} & 0.091 \pm 0.004 {\rm \,(>T6)} &BCS13\\
ULASJ092608.82+040239.7 & - & 19.70 \pm 0.09 & 18.59 \pm 0.06 & - & - & -0.69 \pm 0.14 & T5.9 & T5.3 & T6.4 & T6& 0.317 \pm 0.007 {\rm \,(T6)} & 0.309 \pm 0.0019 {\rm \,(T6)} & 0.266 \pm 0.025 {\rm \,(T4)} &BCS13\\
ULASJ092744.20+341308.7$^\dagger$ & 21.80 \pm 100.00 & 19.66 \pm 0.14 & 18.77 \pm 0.11 & - & - & -1.27 \pm 0.28 & T7.6 & T6.9 & T8.1 & T5.5& 0.347 \pm 0.011 {\rm \,(T6)} & 0.334 \pm 0.022 {\rm \,(T6)} & 0.166 \pm 0.044 {\rm \,(T6/7)} & BCS13\\
WISE J092906.77+040957.9 & - & 17.89 \pm 0.01 & 16.87 \pm 0.01 & 17.24 \pm 0.01 & 17.61 \pm 0.02 & -0.92 \pm 0.07 & T6.7 & T6.5 & T6.9 & T7& 0.276 \pm 0.002 {\rm \,(T7)} & 0.204 \pm 0.004 {\rm \,(T7)} & 0.092 \pm 0.014 {\rm \,(>T6)} &BCS13\\
ULASJ095429.90+062309.6 & - & 17.73 \pm 0.01 & 16.60 \pm 0.01 & 16.87 \pm 0.01 & 17.05 \pm 0.01 & -0.58 \pm 0.09 & T5.4 & T5.1 & T5.8 & T5& 0.370 \pm 0.003 {\rm \,(T5)} & 0.407 \pm 0.005 {\rm \,(T5)} & 0.209 \pm 0.009 {\rm \,(T5)} &BCS13\\
ULASJ102144.87+054446.1$^\dagger$ & - & 18.82 \pm 0.03 & 17.66 \pm 0.02 & 17.96 \pm 0.02 & 17.97 \pm 0.03 & -0.99 \pm 0.26 & T6.9 & T6.1 & T7.5 & T6& 0.331 \pm 0.002 {\rm \,(T6)} & 0.292 \pm 0.005 {\rm \,(T6)} & 0.113 \pm 0.012 {\rm \,(>T6)} &BCS13\\
ULASJ102940.52+093514.6 & - & 18.24 \pm 0.02 & 17.28 \pm 0.01 & 17.63 \pm 0.01 & 17.64 \pm 0.02 & -1.55 \pm 0.17 & T8.2 & T7.9 & T8.4 & T8& 0.182 \pm 0.001 {\rm \,(>T7)} & 0.117 \pm 0.002 {\rm \,(>T7)} & 0.071 \pm 0.010 {\rm \,(>T6)} &BCS13\\
ULASJ104224.20+121206.8 & - & 19.58 \pm 0.09 & 18.52 \pm 0.06 & 18.90 \pm 0.12 & - & -0.93 \pm 0.15 & T6.7 & T6.3 & T7.1 & T7.5& 0.286 \pm 0.007 {\rm \,(T6/7)} & 0.263 \pm 0.013 {\rm \,(T6)} & 0.279 \pm 0.028 {\rm \,(T4)} &BCS13\\
ULASJ104355.37+104803.4 & - & 19.21 \pm 0.03 & 18.23 \pm 0.02 & 18.58 \pm 0.02 & 18.66 \pm 0.05 & -1.36 \pm 0.22 & T7.8 & T7.3 & T8.2 & T8& 0.221 \pm 0.003 {\rm \,(T7)} & 0.173 \pm 0.006 {\rm \,(T7)} & 0.066 \pm 0.009 {\rm \,(>T6)} &BCS13\\
ULASJ105134.32-015449.8 & - & 18.85 \pm 0.03 & 17.75 \pm 0.02 & 18.07 \pm 0.02 & 18.27 \pm 0.04 & -0.56 \pm 0.14 & T5.4 & T4.8 & T5.9 & T6& 0.354 \pm 0.002 {\rm \,(T6)} & 0.301 \pm 0.004 {\rm \,(T6)} & 0.095 \pm 0.013 {\rm \,(>T6)} &BCS13\\
ULASJ105334.64+015719.7 & - & 19.77 \pm 0.10 & 18.50 \pm 0.06 & - & - & -1.10 \pm 0.16 & T7.2 & T6.8 & T7.5 & T6.5& 0.245 \pm 0.029 {\rm \,(T7)} & 0.267 \pm 0.012 {\rm \,(T6)} & 0.130 \pm 0.035 {\rm \,(T6/7)} &BCS13\\
2MASSJ11101001+0116130 & - & 17.21 \pm 0.01 & 16.14 \pm 0.01 & 16.20 \pm 0.01 & 16.09 \pm 0.01 & -0.44 \pm 0.11 & T4.8 & T4.2 & T5.3 & T5.5& 0.152 {\rm \,(T6)} & 0.335 {\rm \,(T5.5)} & 0.175 {\rm \,(T5.5)} &BG2006\\
ULASJ115229.68+035927.3 & - & 18.54 \pm 0.03 & 17.28 \pm 0.02 & 17.70 \pm 0.05 & 17.77 \pm 0.12 & -0.68 \pm 0.06 & T5.9 & T5.7 & T6.1 & T6&0.345 \pm 0.001 {\rm \,(T6)} & 0.287 \pm 0.002 {\rm \,(T6)} & 0.158 \pm 0.009 {\rm \,(T6)} & BCS13\\
ULASJ122343.35-013100.7 & - & 19.71 \pm 0.13 & 18.70 \pm 0.09 & - & - & -0.66 \pm 0.16 & T5.8 & T5.1 & T6.3 & T6&0.315 \pm 0.037 {\rm \,(T6/7)} & 0.434 \pm 0.018 {\rm \,(T5)} & 0.293 \pm 0.058 {\rm \,(T4/5)} & BCS13\\
ULASJ125939.44+293322.4 & - & 19.65 \pm 0.09 & 18.39 \pm 0.06 & 18.55 \pm 0.14 & - & -0.59 \pm 0.13 & T5.5 & T4.9 & T5.9 & T5& 0.408 \pm 0.004 {\rm \,(T5)} & 0.415 \pm 0.006 {\rm \,(T5)} & 0.193 \pm 0.007 {\rm \,(T5)} & BCS13\\
ULASJ130227.54+143428.0 & 19.12 \pm 100.00 & 19.75 \pm 0.13 & 18.60 \pm 0.04 & 18.80 \pm 0.04 & - & -0.23 \pm 0.10 & T3.4 & T2.5 & T4.0 & T4.5&0.432 \pm 0.076 {\rm \,(T5/6)} & 0.516 \pm 0.018 {\rm \,(T4)} & 0.317 \pm 0.054 {\rm \,(T3/4)} & BCS13\\
ULASJ133502.11+150653.5 & - & 19.03 \pm 0.03 & 17.97 \pm 0.02 & 18.30 \pm 0.03 & 18.23 \pm 0.14 & -0.60 \pm 0.09 & T5.6 & T5.2 & T5.9 & T6&0.371 \pm 0.007 {\rm \,(T5)} & - & - & BCS13\\
ULASJ141756.22+133045.8 & 20.38 \pm 0.14 & 17.94 \pm 0.03 & 16.77 \pm 0.01 & 17.00 \pm 0.03 & 17.00 \pm 0.04 & -0.51 \pm 0.08 & T5.1 & T4.7 & T5.4 & T5& 0.381 \pm 0.003 {\rm \,(T5)} & 0.378 \pm 0.002 {\rm \,(T5)} & 0.174 \pm 0.002 {\rm \,(T6)} &BCS13\\
ULASJ142536.35+045132.3 & 21.87 \pm 100.00 & 20.02 \pm 0.14 & 18.70 \pm 0.09 & - & - & -0.93 \pm 0.12 & T6.7 & T6.4 & T7.0 & T6.5& 0.284 \pm 0.008 {\rm \,(T6/7)} & 0.282 \pm 0.012 {\rm \,(T6)} & 0.097 \pm 0.021 {\rm \,(>T6)} &BCS13\\
ULASJ144901.91+114711.4 & - & 18.35 \pm 0.04 & 17.36 \pm 0.02 & 17.73 \pm 0.07 & 18.10 \pm 0.15 & -0.48 \pm 0.13 & T5.0 & T4.3 & T5.5 & T5.5& 0.378 \pm 0.003 {\rm \,(T5)} & 0.389 \pm 0.005 {\rm \,(T5)} & 0.254 \pm 0.010 {\rm \,(T4)} & BCS13\\
ULASJ151637.89+011050.1 & - & 19.48 \pm 0.12 & 18.41 \pm 0.05 & 18.67 \pm 0.06 & 18.49 \pm 0.20 & -0.96 \pm 0.20 & T6.8 & T6.2 & T7.3 & T6.5&0.323 \pm 0.004 {\rm \,(T6)} & - & - & BCS13\\
ULASJ153406.06+055643.9 & 21.77 \pm 0.16 & 20.24 \pm 0.19 & 19.02 \pm 0.10 & - & - & -0.56 \pm 0.13 & T5.4 & T4.8 & T5.8 & T5&0.395 \pm 0.015 {\rm \,(T5)} & 0.442 \pm 0.022 {\rm \,(T5)} & 0.237 \pm 0.021 {\rm \,(T4/5)} & BCS13\\
ULASJ154427.34+081926.6 & - & 19.80 \pm 0.05 & 18.53 \pm 0.03 & 18.49 \pm 0.03 & 18.73 \pm 0.03 & -0.31 \pm 0.16 & T4.0 & T2.8 & T4.8 & T3.5 & 0.520 {\rm \,(T3.5)} & - & - & P2008\\
ULASJ154914.45+262145.6 & - & 19.15 \pm 0.07 & 18.05 \pm 0.03 & 18.29 \pm 0.03 & 18.18 \pm 0.23 & -0.60 \pm 0.12 & T5.5 & T5.0 & T6.0 & T5&0.381 \pm 0.004 {\rm \,(T5)} & - & - & BCS13\\
ULASJ161436.96+244230.1 & 21.31 \pm 0.18 & 19.42 \pm 0.08 & 18.52 \pm 0.04 & - & - & -1.01 \pm 0.15 & T6.9 & T6.6 & T7.3 & T7&0.289 \pm 0.010 {\rm \,(T6/7)} & 0.160 \pm 0.014 {\rm \,(T7/8)} & 0.088 \pm 0.023 {\rm \,(>T6)} & BCS13\\
ULASJ161710.39+235031.4 & - & 18.99 \pm 0.05 & 17.72 \pm 0.02 & 18.16 \pm 0.08 & - & -0.70 \pm 0.09 & T5.9 & T5.6 & T6.2 & T6&0.383 \pm 0.003 {\rm \,(T5)} & 0.348 \pm 0.006 {\rm \,(T6)} & 0.170 \pm 0.007 {\rm \,(T6)} & BCS13\\
ULASJ161934.78+235829.3 & - & 19.72 \pm 0.11 & 18.62 \pm 0.06 & 18.91 \pm 0.06 & 18.41 \pm 0.19 & -0.75 \pm 0.11 & T6.1 & T5.8 & T6.5 & T6&0.273 \pm 0.018 {\rm \,(T6/7)} & - & - & BCS13\\
ULASJ161938.12+300756.4 & - & 19.84 \pm 0.11 & 18.61 \pm 0.07 & 18.79 \pm 0.06 & - & -0.43 \pm 0.09 & T4.7 & T4.3 & T5.1 & T5& 0.441 \pm 0.005 {\rm \,(T5)} & 0.402 \pm 0.010 {\rm \,(T5)} & 0.199 \pm 0.010 {\rm \,(T5)} &BCS13\\
ULASJ211616.26-010124.3 & 22.09 \pm 100.00 & 19.53 \pm 0.12 & 18.27 \pm 0.07 & - & - & -1.10 \pm 0.31 & T7.2 & T6.3 & T7.9 & T6&0.377 \pm 0.011 {\rm \,(T5)} & 0.339 \pm 0.010 {\rm \,(T6)} & 0.138 \pm 0.013 {\rm \,(T6/7)} & BCS13\\
WISE J222623.05+044003.9 & - & 18.04 \pm 0.03 & 16.90 \pm 0.02 & 17.45 \pm 0.07 & 17.24 \pm 0.09 & -1.47 \pm 0.08 & T8.0 & T7.9 & T8.1 & T8&- & - &- & K2011\\
ULASJ230049.08+070338.0 & 21.66 \pm 0.19 & 18.97 \pm 0.04 & 17.67 \pm 0.02 & 17.77 \pm 0.03 & 17.74 \pm 0.05 & -0.43 \pm 0.08 & T4.7 & T4.3 & T5.1 & T4.5&0.504 \pm 0.009 {\rm \,(T4)} & -& - & BCS13\\
ULASJ232600.40+020139.2 & - & 19.40 \pm 0.08 & 17.98 \pm 0.04 & 18.46 \pm 0.12 & 18.41 \pm 0.20 & -1.64 \pm 0.16 & T8.3 & T8.1 & T8.5 & T8& 0.169 \pm 0.012 {\rm \,(>T7)} & 0.097 \pm 0.006 {\rm \,(>T7)} & 0.046 \pm 0.010 {\rm \,(>T6)} &BCS13\\
ULASJ235204.62+124444.9 & - & 19.64 \pm 0.11 & 18.27 \pm 0.05 & 18.55 \pm 0.16 & 18.41 \pm 0.21 & -0.95 \pm 0.12 & T6.8 & T6.5 & T7.1 & T6.5& 0.312 \pm 0.006 {\rm \,(T6)} & 0.206 \pm 0.014 {\rm \,(T7)} & 0.091 \pm 0.021 {\rm \,(>T6)} &BCS13\\

\hline
  \multicolumn{15}{l}{$^\dagger$ These targets had fewer than 10 calibration stars in the field to calibrate the methane photometric zero point.}\\
  \multicolumn{15}{l}{The SpT references column indicates the authors that determined the spectral types and spectral indices listed in the precedent columns, where:  BCS13: \citet{ben2013}, BG2006: \cite{burgasser06}, P2008: \citet{pinfield08}, K2011: \cite{kirkpatrick2011} }\\
\end{tabular}
}
\caption{Photometric and spectroscopic properties of the T~dwarfs analysed in this work that have published spectra. z band photometry was obtained with LRS or SDSS. Y, J, H and K are from the UKIDSS survey and are presented in the MKO system.
The methane photometric type is the conversion of the methane colour using Equation\,\ref{eqn:Tinney2} taking into account photometric errors. 
CH$_4$--J, CH$_4$--H and CH$_4$--K are the methane spectral indices published by the referred authors. 
\label{tab:spec}
}

\end{table}
\end{landscape}

\begin{figure}
  \resizebox{\hsize}{!}{\includegraphics{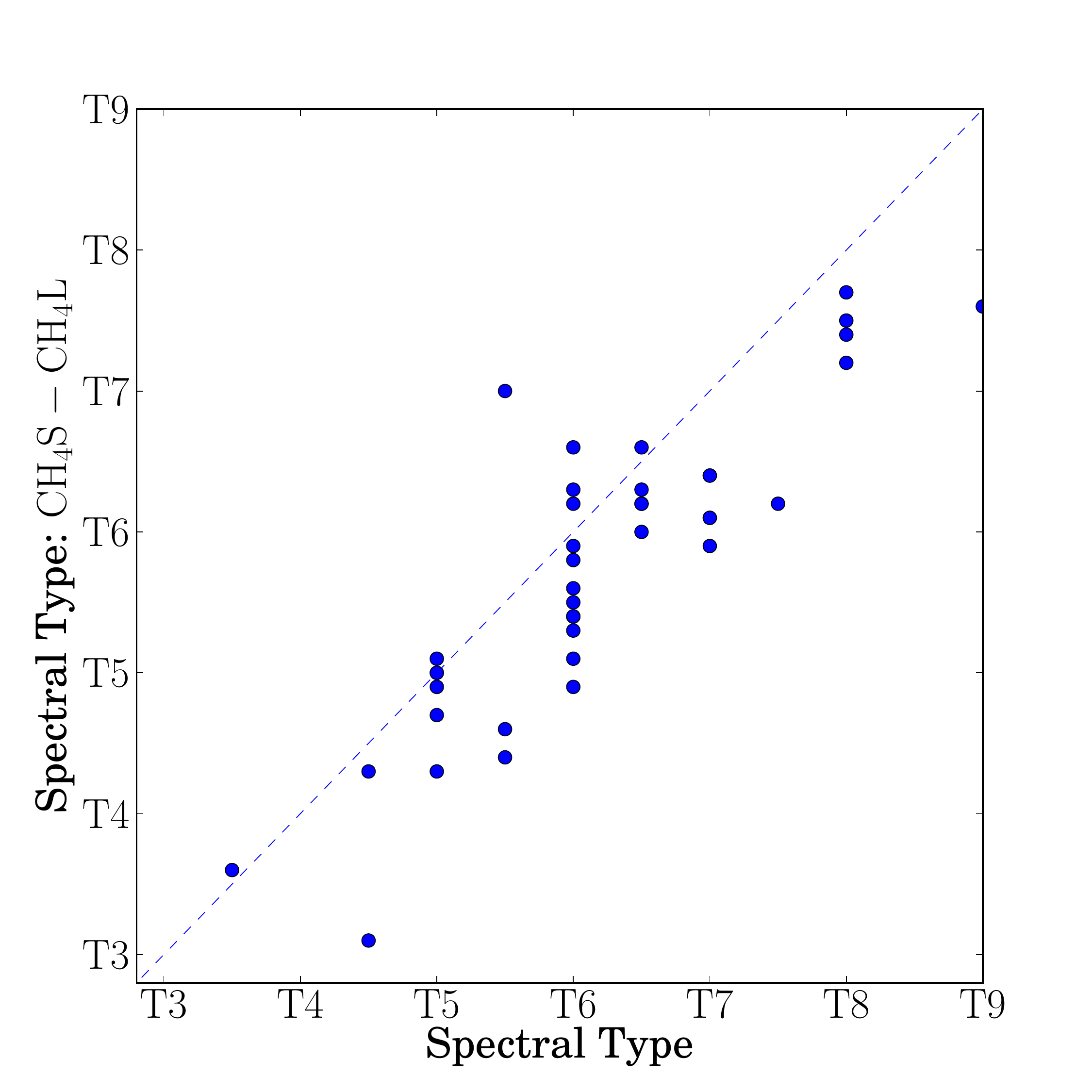}}
 \caption[.]{Comparison between the spectral types of the T~dwarfs of our sample and the spectral types estimated using the $\rm{CH_4}$ colours and Equation\,\ref{eqn:Tinney2}. This plot shows that the spectral types estimated using this method are underestimated.}
\label{fig:TinneyvsSpec}
\end{figure}

 As can be seen by the poor fit in Figure~\ref{fig:TinneyvsSpec}, the later T~dwarfs are classified on average as one spectral subtype earlier using our methane photometry with Equation~\ref{eqn:Tinney2} than when classified by follow-up spectroscopy.

To correct the discrepancy shown in Figure~\ref{fig:TinneyvsSpec} we have recalculated the fit to convert {\chl} -- {\chs} to spectral type for T~dwarfs based in the spectroscopic sample presented in Table~\ref{tab:spec}.
We have used a function with the same shape as the one defined in Equation~\ref{eqn:Tinney2}, and we have recalculated the best fit parameters using a least squares fitting. We have also used the objects with spectral types earlier than T3 presented by \citet{tinney2005} to constrain our fit.
We have refitted our total sample of T~dwarfs ({\chs} - {\chl} $< 0$) with this new defined fit, and the earlier objects with Equation~\ref{eqn:Tinney2}.
Our best fit is given by:

\begin{equation}
\label{eqn:newfit_Tinney2}
\begin{split}
\rm{CH{_4}s} - \rm{CH{_4}l} = & n (0.0067 + 5.900\times10^{-6}n^2 \\
                                               & + 4.172\times10^{-10}n^4 - \dfrac{0.383}{68.5 - n} )
\end{split}
\end{equation}

This fit is only constrained for $n$ between 57.5 and 62 (i.e. T3.5 -- T8). 
The comparison between the fit given by Equation~\ref{eqn:Tinney2} and Equation~\ref{eqn:newfit_Tinney2} is presented in Figure~\ref{fig:comptin}.

\begin{figure}
  \resizebox{\hsize}{!}{\includegraphics{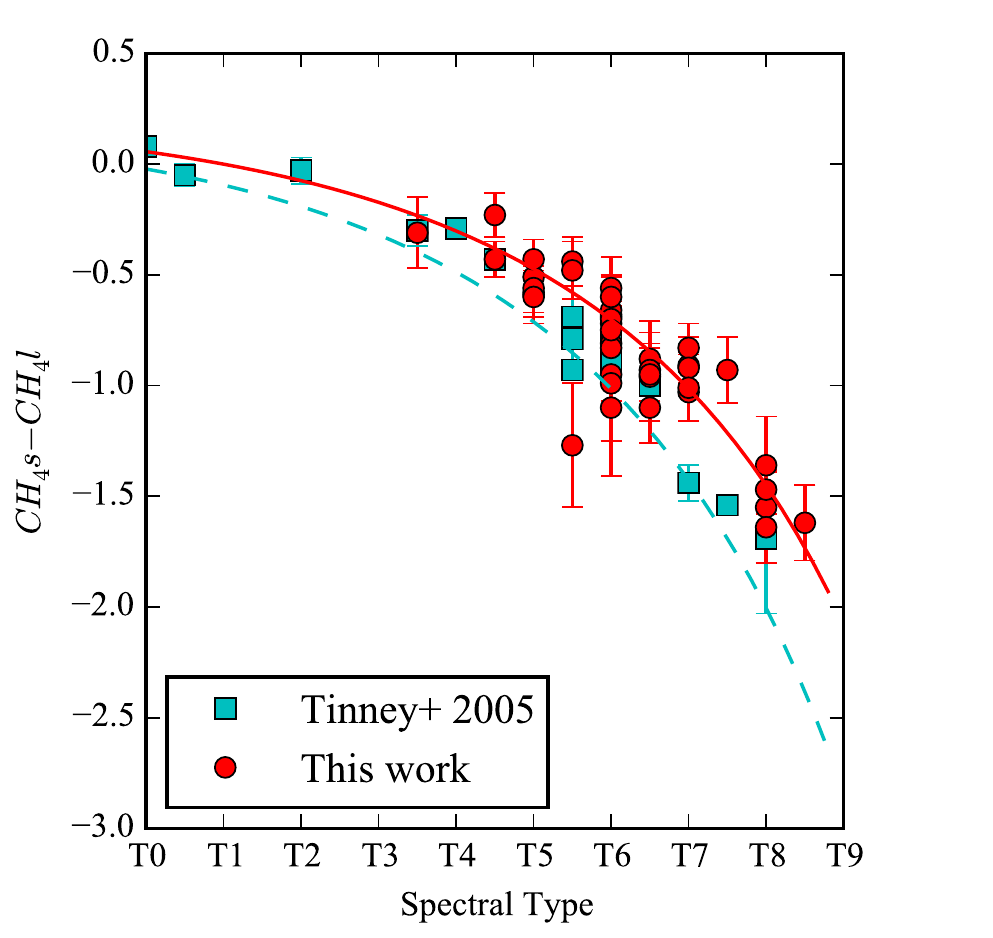}}
 \caption[.]{Comparison between methane colour and spectral type. The cyan squares and dashed line show the data and transformation curve derived by \cite{tinney2005}, whilst the red circles and solid line show the same for this work. Only objects with photometric errors below 0.2 magnitudes were used to calculate the transformation equation.}
\label{fig:comptin}
\end{figure}

Because our methane photometry is exclusively differential in nature (i.e. no standard stars were observed), it is difficult to be certain as to the origin of the discrepancy between the two {\chs} - {\chl} vs. spectral type transforms. 
Although the two sets of methane observations were taken on differing equipment, the filters and detectors have matching specifications. In Figure~\ref{fig:colours} we compare the $J-H$ colours of the objects used to derive the Tinney et al. (2005) transform, and those used here. There is no systematic offset that might suggest a bias in either groups of targets. It should also be noted that the two sets of T~dwarfs have been spectroscopically classified using the same \citet{burgasser06} scheme. 
However, Figure~\ref{fig:comptin} shows that the Tinney et al (2005) fit is constrained for types later than T6.5 by just three objects, and two of these lie significantly below the sources used for our fit. It thus appears that the Tinney et al (2005) transform slightly over predicts the decrease in {\chs} - {\chl} with spectral type as result of the impact of these possible outliers. 

\begin{figure}
  \resizebox{\hsize}{!}{\includegraphics{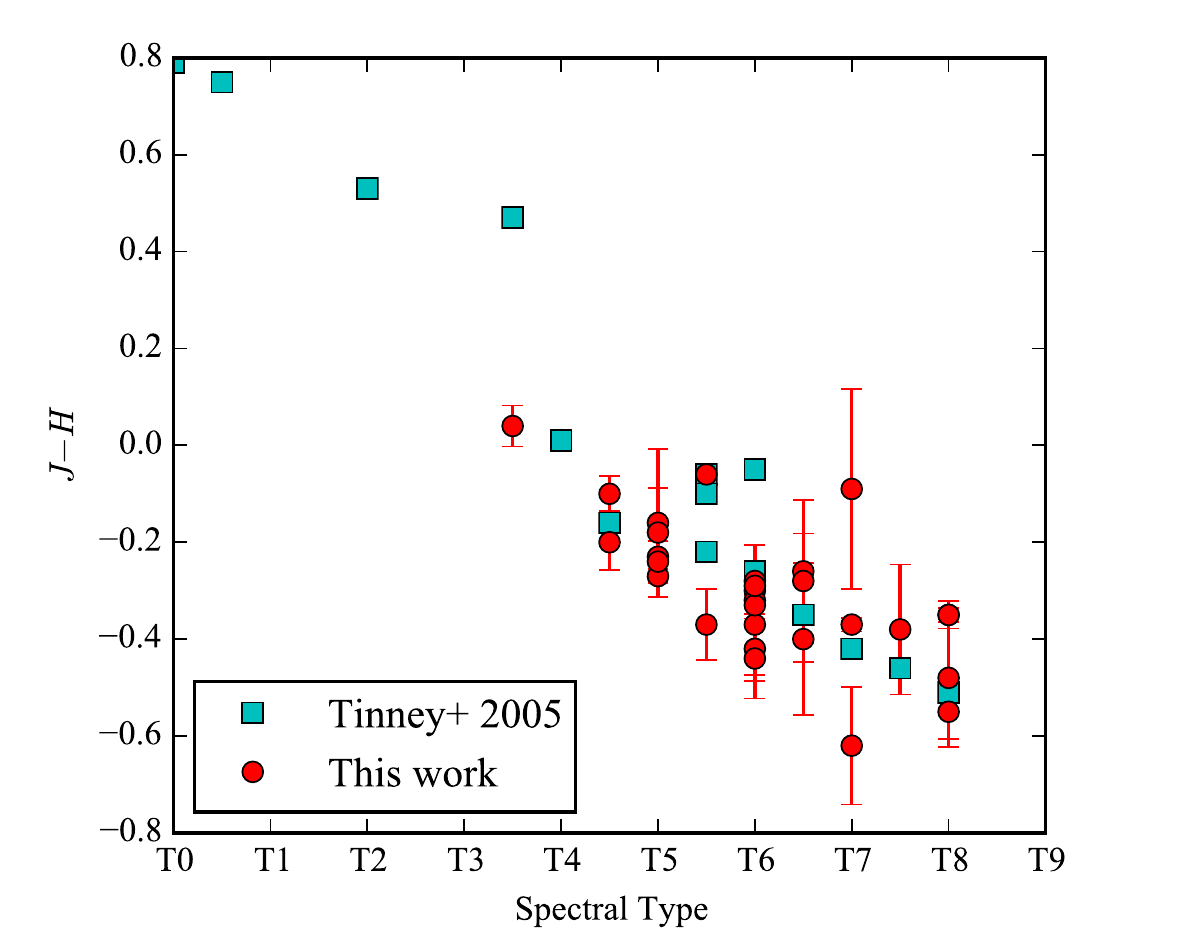}}
 \caption[.]{Comparison between $J-H$ and spectral type for our spectroscopically confirmed T~dwarfs (red circles) and the objects used to define the methane system of \citet[][cyan squares]{tinney2005}.}
\label{fig:colours}
\end{figure}

\subsection{New T dwarfs}

The methane imaging described above was used to shortlist candidates for follow-up spectroscopy in BCS13. All targets classified as T~dwarfs by methane imaging that were observed spectroscopically were confirmed as bona-fide T~dwarfs, with spectral types consistent with those predicted by the method described here.

In addition to the spectroscopically characterised T dwarfs published in BCS13, we present here another 49 objects without follow-up spectroscopy which we classify as T dwarfs. 
For spectral types earlier than T (i.e. stars including M and L~ dwarfs) {\chs} - {\chl}$ \sim 0$, and for early T~dwarfs the gradient of  {\chs} - {\chl} with spectral type is shallow. This means that high signal to noise is required for confident spectral typing of such objects. In this context we have three objects classified between T1 and T3, but whose uncertainties also allow earlier spectral types.

The total sample of photometrically classified T~dwarfs that were not followed up spectroscopically is presented in Table~\ref{tab:phot_T}. We present three estimates of spectral type for each object: a minimum, a maximum and ``best estimate". The ``best" type refers to the spectral type found by inverting Equation~\ref{eqn:newfit_Tinney2} for the measured {\chs} - {\chl}. The maximum and minimum estimates correspond to the types found by inverting Equation~\ref{eqn:newfit_Tinney2} for the 1$\sigma$ error bounds on our measured {\chs} - {\chl}.  About a third of our new T~dwarfs are detected in WISE, and we also present that photometry in Table~\ref{tab:phot_T}.

\begin{landscape}
\begin{table}
\addtocounter{table}{0}
{\scriptsize
\begin{tabular}{ c >{$}c<{$}  >{$}c<{$}  >{$}c<{$}  >{$}c<{$}  >{$}c<{$}  >{$}c<{$}    c  c  c  c c c c c }
\hline
Name & z' & \rm Y_{MKO} & \rm J_{MKO} & \rm H_{MKO} & \rm K_{MKO} &  {\rm CH_4s} - {\rm CH_4l} & \multicolumn{3}{c}{${\rm CH_4}$ Photometric Type}  & W1   &   W2 & W3 & W4 & WISE   \\
&  & & & & & & Best & Min & Max &  &  & & & blend \\
\hline

ULASJ000010.43+113602.2 & 21.61 \pm 100.00 & 19.90 \pm 0.15 & 18.65 \pm 0.08 & -  & -  & -0.04 \pm 0.19 & T1.5 & -- & T3.3 & -- & -- & -- & -- & -- \\
ULASJ002740.95+085940.3 & - & 19.35 \pm 0.09 & 18.20 \pm 0.05 & 18.32 \pm 0.07 & 18.22 \pm 0.15 & -0.30 \pm 0.11 & T3.9 & T3.1 & T4.5 & -- & -- & -- & -- & -- \\
ULASJ004030.42+091524.9 & - & 19.20 \pm 0.08 & 18.16 \pm 0.05 & 18.13 \pm 0.14 & 18.13 \pm 0.21 & -0.37 \pm 0.07 & T4.3 & T3.9 & T4.7 &  $16.88 \pm  0.164$  & $15.92  \pm 0.284$  & $>12.391$ &  $>8.794$  &   Y \\
ULASJ010324.69+092746.5 & 22.29 \pm 0.29 & 19.77 \pm 0.13 & 18.63 \pm 0.08 & -  & -  & -0.75 \pm 0.16 & T6.1 & T5.6 & T6.6 & -- & -- & -- & -- & -- \\
ULASJ010557.68+034643.6 & - & 20.03 \pm 0.14 & 18.61 \pm 0.07 & 18.62 \pm 0.17 & -  & -0.31 \pm 0.12 & T4.0 & T3.0 & T4.7 & -- & -- & -- & -- & -- \\
ULASJ011438.94+085420.8 & 20.94 \pm 100.00 & 19.72 \pm 0.18 & 18.77 \pm 0.12 & -  & -  & -0.62 \pm 0.22 & T5.6 & T4.6 & T6.4 & -- & -- & -- & -- & -- \\
ULASJ014022.04+123756.9 & - & 19.87 \pm 0.11 & 18.48 \pm 0.05 & 18.84 \pm 0.15 & - & -0.61 \pm 0.15 & T5.6 & T4.9 & T6.1 & -- & -- & -- & -- & -- \\
ULASJ014443.27+014741.0 & - & 19.15 \pm 0.08 & 17.94 \pm 0.05 & 18.24 \pm 0.13 & -  & -0.30 \pm 0.09 & T3.9 & T3.3 & T4.4 & $17.916 \pm  0.33$  & $15.889  \pm 0.191$  & $>12.828$  &  $>8.501$  &   N \\
ULASJ020543.26+084340.2 & 22.51 \pm 0.33 & 19.83 \pm 0.20 & 18.67 \pm 0.10 & 19.01 \pm 0.25 & - & -1.52 \pm 0.17 & T8.1 & T7.8 & T8.3  & -- & -- & -- & -- & -- \\
ULASJ021750.35+035803.3$^\dagger$ & - & 20.04 \pm 0.17 & 18.44 \pm 0.07 & 18.80 \pm 0.15 & - & -0.76 \pm 0.10 & T6.2 & T5.8 & T6.4 & -- & -- & -- & -- & -- \\
ULASJ023144.49+063602.2 & - & 20.07 \pm 0.18 & 18.98 \pm 0.11 & -  & -  & -0.17 \pm 0.20 & T2.8 & -- & T4.3 & -- & -- & -- & -- & -- \\
ULASJ032524.61+051039.5 & - & 20.09 \pm 0.16 & 18.97 \pm 0.12 & -  & -  & -0.66 \pm 0.10 & T5.8 & T5.4 & T6.1 & -- & -- & -- & -- & -- \\
ULASJ083408.92+040728.7 & - & 18.92 \pm 0.03 & 17.73 \pm 0.02 & 17.83 \pm 0.02 & 17.94 \pm 0.03 & -0.44 \pm 0.11 & T4.8 & T4.2 & T5.3 & -- & -- & -- & -- & -- \\
ULASJ084743.93+035040.2 & 21.90 \pm 0.10 & 19.61 \pm 0.05 & 18.53 \pm 0.04 & 18.71 \pm 0.03 & 18.99 \pm 0.08 & -0.65 \pm 0.14 & T5.7 & T5.2 & T6.2 & -- & -- & -- & -- & -- \\
ULASJ091804.71+043949.6 & 22.13 \pm 100.00 & 20.06 \pm 0.14 & 18.67 \pm 0.07 & -  & -  & -0.16 \pm 0.21 & T2.8 & -- & T4.3 & -- & -- & -- & -- & -- \\
ULASJ092608.93+020807.3 & 20.93 \pm 100.00 & 19.71 \pm 0.13 & 18.56 \pm 0.07 & -  & -  & -0.48 \pm 0.19 & T5.0 & T3.9 & T5.8 & -- & -- & -- & -- & -- \\
ULASJ110242.39+012021.9 & - & 19.94 \pm 0.11 & 18.43 \pm 0.06 & -  & -  & -0.28 \pm 0.08 & T3.8 & T3.2 & T4.3 & -- & -- & -- & -- & -- \\
ULASJ112059.41+121904.5 & - & 19.20 \pm 0.07 & 17.93 \pm 0.04 & 18.00 \pm 0.05 & 17.94 \pm 0.12 & -0.60 \pm 0.14 & T5.5 & T4.9 & T6.0 & $18.214 \pm  0.44$  & $15.566  \pm 0.147$  & $>12.748$  &  $>9.141$   &  N \\
ULASJ112327.02+150059.4$^\dagger$ & 22.22 \pm 100.00 & 19.61 \pm 0.08 & 18.63 \pm 0.06 & -  & -  & -0.47 \pm 0.14 & T4.9 & T4.2 & T5.6 & -- & -- & -- & -- & -- \\
ULASJ113716.54+112657.2 & 22.24 \pm 100.00 & 20.14 \pm 0.21 & 18.50 \pm 0.09 & -  & -  & -0.51 \pm 0.16 & T5.1 & T4.3 & T5.8 & $16.612 \pm  0.115$  & $15.518  \pm 0.147$  & $>12.073$  &  $>9.01$   &  Y \\
ULASJ114340.47+061358.9 & 20.49 \pm 0.13^{S} & 17.93 \pm 0.02 & 16.82 \pm 0.01 & 16.80 \pm 0.03 & 16.97 \pm 0.07 & -0.36 \pm 0.07 & T4.3 & T3.9 & T4.7 & $16.742 \pm  0.138$  & $14.797  \pm 0.092$  & $>12.464$ &  $>8.763$  &   N \\
ULASJ115533.45-001329.8 & 22.43 \pm -- & 19.84 \pm 0.16 & 18.79 \pm 0.11 & -  & -  & -0.29 \pm 0.19 & T3.8 & T2.1 & T4.9 & -- & -- & -- & -- & -- \\
ULASJ120948.38+035338.4 & 20.51 \pm 0.20 & 19.24 \pm 0.07 & 17.91 \pm 0.04 & 18.07 \pm 0.11 & -  & -0.45 \pm 0.08 & T4.8 & T4.4 & T5.2 & -- & -- & -- & -- & -- \\
ULASJ122633.36+152106.9 & 21.46 \pm 0.21 & 19.81 \pm 0.13 & 18.72 \pm 0.09 & -  & -  & -0.23 \pm 0.11 & T3.4 & T2.5 & T4.2 & -- & -- & -- & -- & -- \\
ULASJ124639.33+032314.3 & - & 18.95 \pm 0.06 & 17.60 \pm 0.03 & 18.05 \pm 0.13 & -  & -0.49 \pm 0.06 & T5.0 & T4.7 & T5.3 & -- & -- & -- & -- & -- \\
ULASJ125015.59+262846.8 & - & 17.75 \pm 0.02 & 16.40 \pm 0.01 & 16.74 \pm 0.02 & 16.79 \pm 0.05 & -0.86 \pm 0.08 & T6.5 & T6.3 & T6.7 & $16.359 \pm  0.087$  & $14.58  \pm 0.065$  & $>12.838$  &  $>9.06$   &  N \\
ULASJ125359.78+265855.5 & - & 19.82 \pm 0.10 & 18.61 \pm 0.06 & 18.75 \pm 0.13 & -  & -0.39 \pm 0.15 & T4.5 & T3.5 & T5.2 & -- & -- & -- & -- & -- \\
ULASJ125446.35+122215.7$^\dagger$ & - & 19.51 \pm 0.11 & 18.29 \pm 0.06 & 18.62 \pm 0.17 & 18.26 \pm 0.20 & -0.56 \pm 0.19 & T5.4 & T4.4 & T6.1 & -- & -- & -- & -- & -- \\
ULASJ130154.79+064747.9 & - & 18.97 \pm 0.07 & 17.79 \pm 0.04 & 18.17 \pm 0.09 & 18.53 \pm 0.26 & -0.72 \pm 0.08 & T6.0 & T5.8 & T6.2 & -- & -- & -- & -- & -- \\
ULASJ130444.26+310112.3 & 22.72 \pm 0.31 & 19.91 \pm 0.14 & 18.79 \pm 0.09 & -  & -  & -1.04 \pm 0.13 & T7.0 & T6.7 & T7.3 & -- & -- & -- & -- & -- \\
ULASJ130716.63+332523.8 & - & 20.01 \pm 0.13 & 18.42 \pm 0.06 & 18.71 \pm 0.16 & -  & -0.49 \pm 0.10 & T5.0 & T4.5 & T5.5 & $13.57 \pm  0.025$  & $13.629  \pm 0.036$  & $>12.707$  &  $>9.208$   &  Y \\
ULASJ131304.03+225919.7 & - & 19.21 \pm 0.05 & 17.91 \pm 0.03 & 18.26 \pm 0.07 & 18.49 \pm 0.18 & -0.64 \pm 0.14 & T5.7 & T5.1 & T6.2 & $18.373 \pm  0.482$  & $15.772  \pm 0.169$  & $>12.231$ &  $>8.752$   &  N \\
ULASJ131313.57+265433.8 & - & 19.64 \pm 0.11 & 18.40 \pm 0.06 & 18.44 \pm 0.10 & 18.66 \pm 0.24 & -0.41 \pm 0.11 & T4.6 & T3.9 & T5.1 & -- & -- & -- & -- & -- \\
ULASJ131610.28+075553.0 & - & 20.00 \pm 0.14 & 19.29 \pm 0.12 & --  & --  & -0.45 \pm 0.13 & T4.8 & T4.1 & T5.4 & $16.491 \pm  0.079$  & $15.356  \pm 0.098$  & $12.239 \pm  0.277$ & $ >9.398 $ &   Y \\
ULASJ131615.93+331306.8 & 21.03 \pm 0.20^{S} & 18.71 \pm 0.04 & 17.49 \pm 0.03 & 17.89 \pm 0.07 & 17.88 \pm 0.14 & -0.64 \pm 0.11 & T5.7 & T5.3 & T6.1 & $17.505 \pm  0.167$  & $15.504  \pm 0.101$  & $13.046 \pm  0.476$ & $ >9.567 $ &   N \\
ULASJ132125.90+073549.4 & - & 19.93 \pm 0.14 & 18.41 \pm 0.08 & 19.14 \pm 0.24 & -  & -0.77 \pm 0.16 & T6.2 & T5.6 & T6.7 & -- & -- & -- & -- & -- \\
ULASJ133750.47+263648.7$^\dagger$ & 20.36 \pm 0.11^{S} & 17.72 \pm 0.02 & 16.56 \pm 0.01 & 16.81 \pm 0.03 & 17.02 \pm 0.06 & -0.51 \pm 0.11 & T5.1 & T4.6 & T5.6 & $16.307 \pm  0.082$  & $14.616  \pm 0.062$  & $>12.224$ & $ >9.032 $  &  N \\
ULASJ134646.71+282009.2$^\dagger$ & - & 18.95 \pm 0.06 & 17.45 \pm 0.03 & 17.59 \pm 0.05 & 17.60 \pm 0.10 & -0.50 \pm 0.08 & T5.1 & T4.7 & T5.4 &  $17.692 \pm  0.234$  & $15.542  \pm 0.125$  & $>12.943$ &  $>9.291$  &   N \\
ULASJ134926.40+234045.9$^\dagger$ & - & 19.87 \pm 0.14 & 18.77 \pm 0.09 & -  & -  & -1.44 \pm 0.25 & T7.9 & T7.5 & T8.3 & -- & -- & -- & -- & -- \\
ULASJ135322.83+283408.2 & - & 19.69 \pm 0.13 & 18.41 \pm 0.07 & 18.92 \pm 0.18 & -  & -1.12 \pm 0.23 & T7.2 & T6.6 & T7.7 &  $14.978 \pm  0.033$  & $14.83  \pm 0.057$  & $12.852 \pm  0.37$ &  $9.265  \pm 0.384$  &   Y \\
ULASJ141520.66+041647.1 & - & 19.19 \pm 0.06 & 18.25 \pm 0.04 & -  & -  & -0.73 \pm 0.12 & T6.1 & T5.7 & T6.4 & -- & -- & -- & -- & -- \\
ULASJ142007.60+021818.3 & - & 19.07 \pm 0.05 & 17.86 \pm 0.03 & 18.09 \pm 0.08 & 18.58 \pm 0.25 & -0.71 \pm 0.08 & T6.0 & T5.7 & T6.2 & -- & -- & -- & -- & -- \\
ULASJ153608.64+030556.5 & 21.39 \pm 0.16 & 20.10 \pm 0.18 & 18.63 \pm 0.09 & -  & -  & -0.24 \pm 0.11 & T3.5 & T2.6 & T4.2 & -- & -- & -- & -- & -- \\
ULASJ211317.05+001840.7 & - & 19.59 \pm 0.11 & 18.29 \pm 0.07 & 18.46 \pm 0.18 & 18.47 \pm 0.24 & -0.41 \pm 0.12 & T4.6 & T3.9 & T5.2 & -- & -- & -- & -- & -- \\
ULASJ213352.64-010343.4 & 22.38 \pm 100.00 & 19.98 \pm 0.15 & 18.80 \pm 0.10 & -  & -  & -0.37 \pm 0.12 & T4.4 & T3.6 & T5.0 &  $17.207 \pm  0.198$  & $16.303  \pm 0.33$  & $>12.133$  & $ >8.497  $ &   N \\
ULASJ223917.13+073416.0 & 22.41 \pm 0.30 & 19.69 \pm 0.07 & 18.88 \pm 0.12 & 18.92 \pm 0.07 & 19.73 \pm 0.18 & -0.60 \pm 0.09 & T5.5 & T5.2 & T5.9 & -- & -- & -- & -- & -- \\
ULASJ225023.52-001605.9 & 22.50 \pm 0.27 & 20.00 \pm 0.21 & 18.97 \pm 0.15 & -  & -  & >-0.62^\ddagger  & -- & -- & -- & -- & -- & -- & -- & -- \\
ULASJ225540.22+061412.9 & 22.15 \pm 0.21 & 19.95 \pm 0.09 & 18.90 \pm 0.09 & 19.10 \pm 0.05 & 19.28 \pm 0.08 & -0.65 \pm 0.09 & T5.8 & T5.4 & T6.0 & -- & -- & -- & -- & -- \\
ULASJ232624.07+050931.6 & 22.02 \pm 0.40 & 19.75 \pm 0.15 & 18.61 \pm 0.10 & 18.61 \pm 0.14 & -  & -0.51 \pm 0.12 & T5.1 & T4.5 & T5.6 & -- & -- & -- & -- & -- \\
\hline
  \multicolumn{15}{l}{$^\dagger$ These targets had fewer than 10 calibration stars in the field to calibrate the methane photometric zero point.}\\
  \multicolumn{15}{l}{$^\ddagger$ ULASJ225023.52--001605.9 is a non detection in CH$_4$l.}\\
\end{tabular}
}
\caption{Photometric properties of the targets with T~dwarf methane photometric types. z band photometry was obtained with LRS or SDSS. Y, J, H and K are from the UKIDSS survey and are presented in the MKO system. W1, W2, W3 and W4 are the WISE bands.
The methane photometric type is the conversion of the methane colour using Equation\,\ref{eqn:Tinney2} taking into account photometric errors. 
\label{tab:phot_T}
}

\end{table}
\end{landscape}

Our sample is divided between objects later than T6, which are the minority (12) since this spectral range was prioritised for spectroscopic follow-up, and the earlier objects between T3 and T6, which represent the bulk of our discoveries (36). There is also a non detection in {\chl}, of ULASJ225023.52Ð001605.9, which we classify as later than T6.

There are a total of 22 objects with methane photometry that were not classified as T dwarfs. These objects are presented in Table~\ref{tab:phot_star}.
Given their colours the majority of these objects are expected to be a mix of late M and L dwarfs. However, post analysis using simbad revealed a subdwarf and peculiar white dwarf, so this selection may have other peculiar objects.

\begin{table*}
{\scriptsize
\begin{tabular}{ c  >{$}c<{$}  >{$}c<{$}  >{$}c<{$}  >{$}c<{$}  >{$}c<{$}  >{$}c<{$}    c  c  c c }
\hline
Name & z' & \rm Y_{MKO} & \rm J_{MKO} & \rm H_{MKO} & \rm K_{MKO} &  {\rm CH_4s} - {\rm CH_4l} & W1   &   W2 & W3 & W4  \\
& & & & & & & &  & & \\
\hline
ULASJ001324.94+064929.6$^\dagger$ & - & 19.42 \pm 0.04 & 19.03 \pm 0.06 & 18.16 \pm 0.02 & 17.39 \pm 0.02 & \rm Not \, detected & -- & -- & -- & -- \\
ULASJ014924.60+065901.9 & 20.79 \pm 0.18 & 19.28 \pm 0.09 & 18.78 \pm 0.11 & - & - & 0.15 \pm 0.12 &-- & -- & -- & -- \\
ULASJ022329.87+032748.5 & - & 18.92 \pm 0.06 & 18.14 \pm 0.05 & - & - & 0.21 \pm 0.07 & -- & -- & -- & --  \\
ULASJ105515.54+081650.6 & 18.89 \pm 0.04^{S} & 18.22 \pm 0.03 & 17.96 \pm 0.04 & 17.91 \pm 0.09 & 17.86 \pm 0.12 & 0.03 \pm 0.09 & -- & -- & -- & -- \\
LSPM J1107+0409N$^{\ddagger}$ & 15.68 \pm 0.01^{S} & 17.54 \pm 0.11 & 17.59 \pm 0.03 & 17.60 \pm 0.05 & - & 0.03 \pm 0.12 & $13.353 \pm  0.027$  & $13.116  \pm 0.033$  & $>11.901$  &  $>8.783$   \\
ULASJ114319.98+125114.3 & 21.96 \pm 3.25^{A} & 19.00 \pm 0.06 & 18.44 \pm 0.06 & 17.97 \pm 0.10 & 17.69 \pm 0.13 & 0.05 \pm 0.10 & $12.17 \pm  0.024$  & $12.218  \pm 0.026$  & $12.226 \pm  0.374$ &  $>9.048$  \\
ULASJ120724.16-004131.2$^\dagger$ & - & 19.49 \pm 0.12 & 18.51 \pm 0.09 & - & - & 0.08 \pm 0.21 & -- & -- & -- & -- \\
ULASJ120936.72+014920.2 & 20.62 \pm 0.16 & 19.71 \pm 0.18 & 18.72 \pm 0.12 & - & - & 0.10 \pm 0.10 & -- & -- & -- & --  \\
ULASJ121901.63+143038.4 & - & 19.28 \pm 0.08 & 18.59 \pm 0.07 & - & - & 0.12 \pm 0.08 & -- & -- & -- & --  \\
SDSS J124739.04+064604.5$^{\star}$ & 18.27 \pm 0.02 & 17.79 \pm 0.03 & 17.60 \pm 0.04 & 17.50 \pm 0.09 & 17.75 \pm 0.15 & 0.18 \pm 0.10 & -- & -- & -- & -- \\
ULASJ125149.84+235653.8 & - & 19.00 \pm 0.06 & 18.65 \pm 0.07 & 18.94 \pm 0.19 & - & 0.04 \pm 0.22 & -- & -- & -- & --  \\
ULASJ131858.09-001632.3 & 20.81 \pm 0.23 & 19.85 \pm 0.12 & 18.76 \pm 0.11 & 18.66 \pm 0.17 & 18.30 \pm 0.21 & 0.21 \pm 0.08 & -- & -- & -- & --  \\
ULASJ135816.34+300539.1$^\dagger$ & - & 19.63 \pm 0.12 & 18.70 \pm 0.09 & 18.64 \pm 0.19 & - & 0.36 \pm 0.13 & -- & -- & -- & --  \\
ULASJ142210.00+003023.7 & - & 19.26 \pm 0.11 & 18.47 \pm 0.09 & - & - & 0.07 \pm 0.18 & $15.97 \pm  0.054$  & $16.049  \pm 0.182$  & $>13.014$  & $ >8.916 $  \\
ULASJ144609.14+020300.3 & 21.23 \pm 0.14 & 19.61 \pm 0.14 & 18.91 \pm 0.13 & - & - & 0.15 \pm 0.07 & -- & -- & -- & --  \\
ULASJ153311.90-010612.9 & - & 19.89 \pm 0.21 & 18.72 \pm 0.13 & - & - & 0.07 \pm 0.12 & -- & -- & -- & -- \\
ULASJ155250.22+013606.6 & 20.51 \pm 0.11 & 19.78 \pm 0.12 & 19.09 \pm 0.12 & - & - & 0.11 \pm 0.09 & -- & -- & -- & -- \\
ULASJ203920.56+002638.3 & - & 19.60 \pm 0.14 & 18.72 \pm 0.09 & - & - & 0.26 \pm 0.13 & -- & -- & -- & --  \\
ULASJ214112.84-010954.6 & 21.59 \pm 0.13 & 20.19 \pm 0.24 & 19.07 \pm 0.15 & - & - & 0.12 \pm 0.08 & -- & -- & -- & -- \\
ULASJ215343.25-001626.0 & 22.38 \pm 100.00 & 19.19 \pm 0.10 & 18.66 \pm 0.11 & - & - & 0.08 \pm 0.08 & --  & -- & -- & -- \\
ULASJ221606.39+032159.2 & - & 19.85 \pm 0.14 & 18.77 \pm 0.09 & - & - & 0.25 \pm 0.08 & -- & -- & -- & -- \\
ULASJ223748.61+052039.6 & - & 19.36 \pm 0.11 & 18.45 \pm 0.08 & - & - & 0.16 \pm 0.11 & -- & -- & -- & -- \\
\hline
  \multicolumn{11}{l}{$^\dagger$ These targets had fewer than 10 calibration stars in the field to calibrate the methane photometric zero point.}\\
  \multicolumn{11}{l}{$^\ddagger$ LSPM J1107+0409N is a sdM5 dwarf binary companion to the white dwarf WD1104+044 \citep{2004ApJ...607..426K,2005AJ....129.1483L}.}\\
   \multicolumn{11}{l}{$^\star$ SDSS J124739.04+064604.5 is a DQ peculiar white dwarf \citep{2005AJ....129.1483L,2010ApJS..190...77K}.} \\
\end{tabular}
}
\caption{Photometric properties of the targets classified as non T~dwarfs. Due to their large proper motion a couple of known nearby white dwarfs got scattered into our sample. $z'$ band photometry was obtained with LRS or SDSS. Y, J, H and K bands are from the UKIDSS survey and are presented in the MKO system. W1, W2, W3 and W4 are the WISE bands, that for these objects are all blended detections.
\label{tab:phot_star}
}

\end{table*}

\section[]{Search for wide binary companions}
\label{sec:binary}

To search for wide binary companions we have followed the method of BCS13. We have compared the new  sample of photometric T~dwarfs to Hipparcos, NOMAD, LSPM and the UKIDSS LAS proper motion catalogue  \citep{2007A&A...474..653V,2005AJ....129.1483L,2004AAS...205.4815Z,smith2014}, allowing for upto a 20000~AU projected separation and requiring that candidate pairs have consistent distance estimates. The somewhat larger uncertainties in in the spectral types for our targets, compared to
spectroscopically confirmed objects, means that the possible distance
ranges for our T dwarfs are relatively larger than for BCS13. The angular separation limit for each pair was driven by our 20000~AU projected separation limit and the minimum distance estimate for the T~dwarf.

For the LSPM and NOMAD comparisons we restricted our search to stars with well-measured $V-J$ colours to allow photometric distance estimates for the candidate primary stars, as in BCS13.  The $V-J$ colour was used to place the candidate primary stars on a colour-magnitude diagram, with the absolute magnitude of the ``primary" star estimated from its 2MASS $J$ band apparent magnitude and the maximum and minimum plausible distances to the T~dwarf ``companion". Only candidate binary systems for which the ``primary" staddled the main-sequence defined by Hipparcos stars were retained for further consideration \citep[see ][ for a more detailed example of this method]{ben2013}. Those sources that passed this test were then checked for common motion. The requirement for a good $V-J$ colour (in practice this is mainly limited by the availability of $V$ band data) introduces a significant element of incompleteness in our search for primary stars, particularly for our most distant T~dwarfs. Unfortunately, obtaining a more complete list of potential primary stars is beyond the scope of this work, and the companion search is presented with this caveat in-mind.

Some of our targets already have proper-motions measured by \citet{smith2014}, however in most cases no UKIDSS proper motion was available. In some cases,  two epochs of UKIDSS data were available due to long gaps between the $J$ and $H$ band images, and in these cases a proper motion was estimated using these data, following the same method as in \citet{smith2014}.  In the remainder of cases proper motions were estimated using the UKIDSS survey data for the first epoch, and our follow-up imaging (typically the {\chs} detection) for the second epoch.

Catalogues of sources in our follow-up images were made using \textsc{IMCORE}, as used for our photometry (see Section~\ref{sec:ch4cal}). These catalogues were used to derive a spatial transformation  between from the NICS follow-up coordinate system to the UKIDSS system using the \textsc{GEOMAP IRAF} package.   Since NICS follow-up catalogues had a crude astrometric calibration, the transforms to the well-calibrated UKIDSS system were allowed to take account of non-linear shifts, rotation and scaling (a ``general" fit in \textsc{GEOMAP}). Sources with large residuals to the fit ($>2 \sigma$) were excluded.  We then transformed the NICS coordinates of the targets to the UKIDSS system using \textsc{GEOXYTRAN}, and determined their motion between the epochs. 

In Table~\ref{tab:kinematics} we present the proper motions for all targets we were able to measure. The J, {\chs} measurements presented are relative proper motions, due to small field of view of the NICS field, and the relatively small number of calibration stars. \cite{smith2014} shows that the correction from relative to absolute proper motions is usually of the order of 5~mas/yr and less than 10~mas/yr.
Candidate binary systems were identified for further consideration if their proper motions agreed within $4 \sigma$.

\begin{table*}
\centering
{\scriptsize
\begin{tabular}{ c >{$}c<{$}  >{$}c<{$}  >{$}c<{$}  >{$}c<{$}  >{$}c<{$}  >{$}c<{$} }
\hline
Name &  \rm{\mu_{\alpha}\,cos\,\delta}  & \rm{\mu_{\delta}} & \rm{Epoch \, Baseline} &  \rm{Reference} & \multicolumn{2}{c}{\rm{Photometric \, distance}}  \\
 &  \rm{(mas/yr)}  & \rm{(mas/yr)} & \rm{(years)} &  & \rm{Min \, (pc)} & \rm{Max \, (pc)}  \\

\hline
ULASJ000010.43+113602.20 &  -89 \pm  31 &  -8 \pm 27 &  2.367 & 4 & 57.0  &  - \\
ULASJ002740.95+085940.30 &  205 \pm  17 &  -287 \pm 29 &  1.386 & 4 & 42.0  &  96.0 \\
ULASJ004030.42+091524.90 &  -57 \pm 32 &  -92 \pm 49 & 1.231  & 4 &  47.0 & 94.0 \\
ULASJ010324.69+092746.50 & -155 \pm 27 & -89 \pm  30 & 1.378 & 4 & 40.1  & 103.0 \\
ULASJ010557.68+034643.60 & -70 \pm 37  &  -5 \pm 35 & 1.124  & 4 & 50.0 & 113.3  \\
ULASJ011438.94+085420.80 &  181 \pm 28 &  -155 \pm 28 & 2.301  & 4 & 42.1  & 139.1  \\
ULASJ014022.04+123756.90 &  23 \pm 46  & -113 \pm 47 &  0.759 & 3 \, \rm{rel} & 40.8  & 108.1  \\
ULASJ014443.27+014741.00 &  -140 \pm  41 & -203  \pm 51  & 1.165 & 4 & 37.2  & 85.2 \\
ULASJ020543.26+084340.20 &  291 \pm  66 &  81 \pm 64 & 1.262 & 3 \, \rm{abs} & 12.1  & 66.6  \\
ULASJ021750.35+035803.30 &  368 \pm  74 &  -214 \pm 124 & 0.920 & 4 & 36.9  & 93.8 \\
ULASJ023144.49+063602.20 &  37 \pm 28 &  61 \pm 34  & 2.107 & 4 & 63.9   &  - \\
ULASJ032524.61+051039.50 &  -24 \pm 20  & -3 \pm 19 & 2.128  & 4 & 46.2 & 122.3  \\
ULASJ083408.92+040728.70 &  97 \pm 33  & -330 \pm 34  & 0.971 & 2 & 34.2 & 76.2  \\
ULASJ084743.93+035040.20 &  -159 \pm 25 & 58 \pm 25 & 0.972  & 2 & 42.1  & 110.2  \\
ULASJ091804.71+043949.60 &  98 \pm 24  & -42  \pm 21 & 2.402 & 4 & 56.5 & - \\
ULASJ092608.93+020807.30 &  163 \pm 9  &  -155 \pm 9  &  3.877 & 1 & 42.1 & 114.3  \\
ULASJ110242.39+012021.90$^\dagger$ & -30  \pm 26  &  -36 \pm 28  & 2.349 & 4 & 46.3  & 107.4 \\
ULASJ112059.41+121904.50 & -57 \pm  17 & 1 \pm 19  & 1.934  & 4 &  31.9 & 83.3 \\
ULASJ112327.02+150059.40$^\dagger$ & -118 \pm 25 &  -548 \pm 26  & 1.976  & 4 & 50.8 & 117.3 \\
ULASJ113716.54+112657.20 & 37  \pm 45  & -167  \pm 32  & 1.272 & 4 & 40.6 & 120.8 \\
ULASJ114340.47+061358.90 & -67.2 \pm 13.9  & 30.5  \pm 13.9  & 1.581 & 3 \, \rm{abs}  & 25.8 & 50.0 \\
ULASJ115533.45-001329.80 & -83  \pm 11 & -11 \pm 14 & 4.777 & 4 & 53.5 & 120.8 \\
ULASJ120948.38+035338.40 & -22 \pm 8 & -63  \pm 8 & 4.759 & 1 & 36.9 & 90.0  \\
ULASJ122633.36+152106.90 & -171  \pm 14  & -25  \pm  14 & 3.607  & 1 & 58.6 & 120.2  \\
ULASJ124639.33+032314.30 &  -82 \pm 23  & -37 \pm 17 & 1.864 & 4 & 32.0  & 77.6  \\
ULASJ125015.59+262846.80 & -456 \pm  18 & -580 \pm 16  & 1.869  & 4 & 13.2  & 26.2  \\
ULASJ125359.78+265855.50 & 131  \pm 16  & -48  \pm 15 &  2.243 & 4 & 57.6  & 116.3 \\
ULASJ125446.35+122215.70 & 25  \pm  11 & 28  \pm 11 &  3.803 & 1 & 37.4  & 108.4  \\
ULASJ130154.79+064747.90 & -40  \pm  28 &  -83 \pm 24  & 1.809  & 4 & 27.8 & 68.7 \\
ULASJ130444.26+310112.30 & 3  \pm 17  & 30 \pm 17 & 2.786 & 4 & 25.2  & 87.9  \\
ULASJ130716.63+332523.80 & -63 \pm  26 &  -8 \pm 22 & 1.759 & 4 & 46.1  &  115.2 \\
ULASJ131304.03+225919.70 & 93 \pm 24 &  -163 \pm 22  & 1.935 & 4 &  31.7 & 82.4 \\
ULASJ131313.57+265433.80 &  178 \pm 23 & -141 \pm 18 & 1.924 & 4 & 52.3  & 105.8 \\
ULASJ131610.28+075553.00 & -1003 \pm 13 & 106 \pm 12  & 5.668 & 1 & 67.1 & 164.0 \\
ULASJ131615.93+331306.80 & -48 \pm 44  & -167 \pm 52  & 1.754  & 4 & 26.2 & 67.8 \\
ULASJ132125.90+073549.40 & -40 \pm 19 & 80 \pm 19 & 2.705 & 1 & 36.3 &  93.0 \\
ULASJ133750.47+263648.70 &  -3 \pm 13 & -18 \pm 26  & 1.814 & 4 & 20.0 & 47.7 \\
ULASJ134646.71+282009.20 & -454  \pm 41 & -61 \pm 21 & 1.798 & 4 & 29.8 & 72.5 \\
ULASJ134926.40+234045.90$^\dagger$ & 105 \pm 21 &  -70 \pm  21 &  2.172 & 4 & 12.8 & 69.3 \\
ULASJ135322.83+283408.20 &  42 \pm 27  & -32 \pm 20 & 1.800 & 4 & 21.3 & 73.1 \\
ULASJ141520.66+041647.10 & -84 \pm 21 &  3 \pm 26 & 1.940 & 4 & 34.4  & 84.9 \\
ULASJ142007.60+021818.30 & -59 \pm 28  & -116 \pm 20 & 1.940 & 4  & 28.8  & 70.3  \\
ULASJ153608.64+030556.50 &  -62 \pm 18 & 2 \pm 18 & 2.946 & 1 & 56.2 & 115.3 \\
ULASJ211317.05+001840.70 &  60 \pm 27 & -44 \pm  29 & 1.471  & 4 & 49.5 &  100.9 \\
ULASJ213352.64-010343.40 & -1  \pm 19  &  -104 \pm 25 & 2.393 & 4 & 61.7  & 129.4  \\
ULASJ223917.13+073416.00 & 94  \pm  16 & -74 \pm 16 & 2.140 & 2 & 47.6  &  134.3 \\
ULASJ225540.22+061412.90 &  67 \pm  16 &  40 \pm 17 & 2.146  & 2 & 45.3  &  116.9 \\
ULASJ232624.07+050931.60$^\dagger$ & 0 \pm 21  & -105  \pm 41  & 2.394  & 4 & 49.5 &  127.5 \\
\hline
  \multicolumn{7}{l}{$^\dagger$ For these objects the small number of background astrometric calibration stars ($<15$) may affect the quality of the fit.}\\
\end{tabular}
}

\caption{Kinematic properties of the methane colour selected T dwarfs presented in Table~\ref{tab:phot_T}. 
The reference column refers to the origin of the proper motion measurement: 1 - Proper motion from \citet{smith2014}.
2 - Proper motion calculated using the same method as \citet{smith2014}: the first epoch is a J band UKIDSS measurement and the second epoch is a WFCAM measurement taken in the same conditions as the original UKIDSS data; the values are absolute proper motions.
3 - Proper motion calculated using the same method as \citet{smith2014}: the two epochs are a J and H band UKIDSS measurements;  the values are either relative proper motions or absolute proper motions, denoted by suffix `rel' or `abs' respectively.
4 - Proper motion calculated using the imaging in this work. The first epoch is a J band UKIDSS measurement and the second epoch is the CH$_4$s measurement; the values presented are relative proper motions.
All photometric distances were calculated based on the scale of absolute magnitudes for T~dwarfs by \citet{2012ApJS..201...19D}.
\label{tab:kinematics}
}

\end{table*}

Only one possible common proper motion wide binary system was identified in our sample by this method. ULASJ134926.40+234045.90 (hereafter ULASJ1349+2340), a T7.5--T8.3 dwarf, lies an angular separation of 23~arcminutes from the K4V star HD~120751. HD~120751 has a well measured Hipparcos distance of  $43\pm4$~pc, whilst ULASJ1349+2340 has an estimated distance in the range 12--70~pc. Although the angular separation of the pair passed our projected separation $< 20000$~AU test for the minimum distance to the T~dwarf, the well measured Hipparcos distance to the primary takes priority in this case. If the pair were a genuine common proper motion binary, the projected separation would be $\approx 60000$~AU, well beyond our maximum projected separation cut. 
We have estimated the probability that the pair represent a chance alignment following the method of \citet{dhital2010}. This method calculates the frequency of unrelated pairings using a Galactic model that is parameterised by empirically measured stellar number density \citep{juric2008,bochanski2010} and space velocity \citep{bochanski2007} distributions. All stars in the model are single (and hence unrelated); therefore any stars within the 5D ellipsoid defined by the binary's position, angular separation, distance, and proper motions is a chance alignment. We performed 10$^6$ Monte Carlo realisations to calculate the probability of chance alignment.  This result is a high probability that the pair are a chance alignment ($78\%$). This is partly driven by the  large distance uncertainty for the T~dwarf.  We thus do not rule out the possibility that the pair are a genuine extremely wide binary system, but a trigonometric parallax measurement for ULAS~J1349+2340 would be required to test this further.  

Given the $> 7 \%$ wide companion rate estimated in BCS13 and $> 5 \%$ found by \citet{gomes2013}, we would expect to find $\sim$3 wide companions amongst our set of new T~ dwarfs. A zero find is not inconsistent with previous estimates of the lower limit on the wide companion frequency, but may also reflect incompleteness in the catalogues searched for primary stars. The objects identified in this work sample lie at larger distances than the later type objects that were the focus of BCS13, or the 2MASS detected objects in \citet{gomes2013}, and consequently will have smaller proper motions, and the potential primary stars may be absent or poorly characterised in the catalogues we have searched. We thus note that there may be further wide companions existing within the set of objects presented here, but it is not possible to identify them with the data in hand.

We note that the Gaia mission will do much to improve the selection of benchmark systems by addressing the incompleteness of the primary star catalogues. Its complete dataset will include proper motions and  parallaxes with sub-milliarcsecond precision for stars down to 20th magnitude in the optical \citep{gaia}. As such, the sample of brown dwarfs that have already been searched for benchmark systems should be re-examined in the light of these new data.

\section[]{Concluding remarks}
\label{sec:summary}

We have presented the full details of our methane imaging program that formed a key element of our strategy in building the UKIDSS late-T dwarf sample presented in BCS13. A total of 116 candidate T~dwarfs were observed with methane photometry. Of these, 94 were found to have methane colours consistent with T~dwarf classification. A subset of 45 of these objects have been observed spectroscopically, all of which are confirmed as T~dwarfs, with spectral types consistent with those predicted from their {\chs - \chl} colours (see BCS13 and references therein). The remaining 49 photometrically classifed T~dwarfs are presented here as new discoveries. Given the 100\% success rate in classifying the spectroscopic subset, we consider these confirmed T~dwarfs (within the quoted uncertainties). 
This set of T~dwarfs represents a useful addition to the known population of cool brown dwarfs, and demonstrates that methane imaging continues to provide a low-cost method for characterising photometrically selected T~dwarfs in deep and wide field surveys.

Whilst methane imaging is an excellent follow-up tool, it would not be desirable as an inherent part of wide field surveys themselves given its necessarily niche benefit and the significant additional observing time required to achieve useful depth across any of the current or future surveys discussed in the introduction.  It is more desirable to achieve better complementarity of survey depths across the full red-optical to mid-IR range wavelength range to allow direct characterisation from survey data. As we highlighted in Section~\ref{sec:candsel}, the poor match in depth between UKIDSS and WISE for earlier type T~dwarfs prevented us from effectively using this outstanding dataset as part of our selection strategy.
Ongoing observations by the revived WISE spacecraft as part of the NEOWISE-R mission \citep{neowiseR}, however, provide the opportunity to improve on the depth of the WISE and ALLWISE data sets. This would add significant value to current and future wide field surveys such as VHS, UHS and DES, and this represents the most significant near-term opportunity in substellar survey science.

\appendix
\section{Summary of photometric observations}

\begin{table*}
\begin{tabular}{ c  c  c  c c c c c }
\hline
Object & Filter & Instrument & Date & TNG Program & T int (s) & N coadds & N dither \\
\hline
ULASJ000010.43+113602.2 & CH$_4$s, CH$_4$l  & NICS          & 2012-01-16 & A24 TAC49 & 30  & 4 & 10 \\
                                              & z                               & DOLORES & 2012-01-13 & A24 TAC49 & 450 & 2 & -- \\
ULASJ000734.90+011247.1& CH$_4$s, CH$_4$l  & NICS          & 2011-10-27 & A24 TAC49 & 30  & 1 & 30 \\
ULASJ001324.94+064929.6 & CH$_4$s, CH$_4$l  & NICS          & 2010-12-25 & A22 TAC96 & 30  & 1 & 30 \\
ULASJ002740.95+085940.3 & CH$_4$s, CH$_4$l  & NICS          & 2011-11-18 & A24 TAC49 & 30  & 1 & 30 \\
ULASJ004030.42+091524.9 & CH$_4$s, CH$_4$l  & NICS          & 2011-10-28 & A22 TAC49 & 30  & 1 & 30 \\
ULASJ010324.69+092746.5 & CH$_4$s, CH$_4$l  & NICS          & 2012-01-15 & A24 TAC49 & 30  & 4 & 10 \\
                                              & z                               & DOLORES & 2012-01-14 & A24 TAC49 & 450 & 2 & -- \\
ULASJ010557.68+034643.6 & CH$_4$s, CH$_4$l  & NICS          & 2011-10-28 & A24 TAC49 & 30  & 1 & 30 \\
ULASJ011438.94+085420.8 & CH$_4$s, CH$_4$l  & NICS          & 2013-01-25 & A26 TAC68 & 26 , 30  & 3 , 4 & 30 \\
                                              & z                               & DOLORES & 2012-01-13 & A24 TAC49 & 450 & 2 & -- \\
ULASJ012735.66+153905.9 & CH$_4$s, CH$_4$l  & NICS          & 2011-11-19 & A24 TAC49 & 20  & 2 & 30 \\
ULASJ012855.07+063357.0 & CH$_4$s, CH$_4$l  & NICS          & 2010-11-06 & A22 TAC96 & 20  & 2 & 30 \\
ULASJ013017.79+080453.9 & CH$_4$s, CH$_4$l  & NICS          & 2010-12-25 & A22 TAC96 & 30  & 1 & 30 \\
ULASJ013950.51+150307.6 & CH$_4$s, CH$_4$l  & NICS          & 2011-10-27 & A24 TAC49 & 45  & 1 & 30 \\
ULASJ014022.04+123756.9 & CH$_4$s, CH$_4$l  & NICS          & 2012-01-15 & A24 TAC49 & 30  & 4 & 10 \\
ULASJ014443.27+014741.0 & CH$_4$s, CH$_4$l  & NICS          & 2011-11-19 & A24 TAC49 & 20  & 2 & 30 \\
ULASJ014924.60+065901.9 & CH$_4$s, CH$_4$l  & NICS          & 2012-01-16 & A24 TAC49 & 30  & 4 & 10 \\
                                              & z                               & DOLORES & 2012-01-13 & A24 TAC49 & 450 & 2 & -- \\
ULASJ020013.18+090835.2 & CH$_4$s, CH$_4$l  & NICS          & 2011-10-28 & A24 TAC49 & 30  & 1 & 30 \\
ULASJ020543.26+084340.2 & CH$_4$s, CH$_4$l  & NICS          & 2012-09-22 & A26 TAC68 & 26, 30  & 3, 4 & 30 \\
                                              & z                               & DOLORES & 2012-08-07 & A26 TAC68 & 600 & 2 & -- \\
ULASJ021750.35+035803.3 & CH$_4$s, CH$_4$l  & NICS          & 2012-08-07 & A26 TAC68 & 30  & 4 & 30 \\
ULASJ022329.87+032748.5 & CH$_4$s, CH$_4$l  & NICS          & 2012-09-24 & A26 TAC68 & 26, 30  & 3, 4 & 30 \\
ULASJ023144.49+063602.2 & CH$_4$s, CH$_4$l  & NICS          & 2010-12-25 & A22 TAC96 &  30  & 2 & 30 \\
ULASJ032524.61+051039.5 & CH$_4$s, CH$_4$l  & NICS          & 2010-12-26 & A22 TAC96 &  30  & 2 & 30 \\
ULASJ074502.79+233240.3 & CH$_4$s, CH$_4$l  & NICS          & 2011-10-28 & A24 TAC49 &  60  & 1 & 30 \\
ULASJ075937.75+185555.0 & CH$_4$s, CH$_4$l  & NICS          & 2011-10-27 & A24 TAC49 &  60  & 1 & 30 \\
                                              & z                               & DOLORES & &  &  &  & -- \\
ULASJ081110.86+252931.8 & CH$_4$s, CH$_4$l  & NICS          & 2010-12-27 & A22 TAC96 &  30  & 1 & 30 \\
ULASJ083408.92+040728.7 & CH$_4$s, CH$_4$l  & NICS          & 2011-05-06 & A23 TAC28 &  30  & 1 & 30 \\
ULASJ084743.93+035040.2 & CH$_4$s, CH$_4$l  & NICS          & 2011-05-07 & A23 TAC28 &  30  & 1 & 30 \\
ULASJ091804.71+043949.6 & CH$_4$s, CH$_4$l  & NICS          & 2012-04-28 & A25 TAC32 &  30  & 4 & 10 \\
ULASJ092608.82+040239.7 & CH$_4$s, CH$_4$l  & NICS          & 2012-04-29 & A25 TAC32 &  30  & 4 & 10 \\
ULASJ092608.93+020807.3 & CH$_4$s, CH$_4$l  & NICS          & 2012-04-28 & A25 TAC32 &  30  & 4 & 10 \\
ULASJ092744.20+341308.7 & CH$_4$s, CH$_4$l  & NICS          & 2011-05-12 & A23 TAC28 &  30  & 2 & 30 \\
                                              & z                               & DOLORES & &  &  &  & -- \\
WISE J092906.77+040957. & CH$_4$s, CH$_4$l  & NICS          & 2011-05-07 & A23 TAC28 &  30  & 1 & 30 \\
ULASJ095429.90+062309.6 & CH$_4$s, CH$_4$l  & NICS          & 2011-05-09 & A23 TAC28 &  30  & 1 & 30 \\
ULASJ102144.87+054446.1 & CH$_4$s, CH$_4$l  & NICS          & 2012-01-16 & A24 TAC49 &  30  & 4 & 10 \\
ULASJ102940.52+093514.6 & CH$_4$s, CH$_4$l  & NICS          & 2011-05-09 & A23 TAC28 &  30  & 1 & 30 \\
ULASJ104224.20+121206.8 & CH$_4$s, CH$_4$l  & NICS          & 2012-01-16 & A24 TAC49 &  30  & 4 & 10 \\
ULASJ104355.37+104803.4  & CH$_4$s, CH$_4$l  & NICS          & 2011-05-09 & A23 TAC28 &  20  & 2 & 30 \\
ULASJ105134.32-015449.8  & CH$_4$s, CH$_4$l  & NICS          & 2011-05-10 & A23 TAC28 &  30  & 1 & 30 \\
ULASJ105334.64+015719.7 & CH$_4$s, CH$_4$l  & NICS          & 2012-01-16 & A24 TAC49 &  30  & 4 & 10 \\
ULASJ105515.54+081650.6  & CH$_4$s, CH$_4$l  & NICS         & 2011-05-10 & A23 TAC28 &  30  & 1 & 30 \\
                                              & z                               & DOLORES & &  &  &  & -- \\
ULASJ110242.39+012021.9 & CH$_4$s, CH$_4$l  & NICS         & 2012-04-29 & A25 TAC32 &  30  & 4 & 10 \\
LSPM J1107+0409N & CH$_4$s, CH$_4$l  & NICS         & 2011-06-01 & A23 TAC28 &  30  & 1 & 30 \\
2MASSJ11101001+0116130  & CH$_4$s, CH$_4$l  & NICS         & 2012-01-15 & A24 TAC49 &  20  & 3 & 10 \\
ULASJ112059.41+121904.5 & CH$_4$s, CH$_4$l  & NICS         & 2012-01-17 & A24 TAC49 &  30  & 4 & 10 \\
ULASJ112327.02+150059.4 & CH$_4$s, CH$_4$l  & NICS         & 2012-02-04 & A25 TAC32 &  30  & 4 & 10 \\
                                              & z                               & DOLORES & 2012-01-25 & A24 TAC49 & 600 & 2 & -- \\
ULASJ113115.64 +054312.4 & CH$_4$s, CH$_4$l  & NICS         & 2011-05-11 & A23 TAC28 &  30  & 1 & 30 \\                                              
ULASJ113716.54+112657.2 & CH$_4$s, CH$_4$l  & NICS         & 2011-05-10 & A23 TAC28 &  20  & 2 & 30 \\
                                              & z                               & DOLORES & &  &  &  & -- \\
ULASJ114319.98+125114.3 & CH$_4$s, CH$_4$l  & NICS         & 2010-12-26 & A22 TAC96 &  30  & 1 & 30 \\
                                              & z                               & ACAM & &  &  &  & -- \\
ULASJ114340.47+061358.9 & CH$_4$s, CH$_4$l  & NICS         & 2013-01-25 & A26 TAC68 &  30  & 4 & 10 \\
ULASJ115229.68+035927.3 & CH$_4$s, CH$_4$l  & NICS         & 2010-12-26 & A22 TAC96 &  30  & 1 & 30 \\
ULASJ115533.45-001329.8 & CH$_4$s, CH$_4$l  & NICS         & 2012-02-01 & A24 TAC49 &  30  & 4 & 10 \\
ULASJ120724.16-004131.2 & CH$_4$s, CH$_4$l  & NICS         & 2012-04-28 & A25 TAC32 &  30  & 1 & 10 \\
ULASJ120936.72+014920.2 & CH$_4$s, CH$_4$l  & NICS         & 2013-01-25 & A26 TAC68 &  26, 30  & 3, 4 & 30 \\
ULASJ120948.38+035338.4 & CH$_4$s, CH$_4$l  & NICS         & 2013-01-25 & A26 TAC68 &  30  & 2 & 30 \\
ULASJ121901.63+143038.4 & CH$_4$s, CH$_4$l  & NICS         & 2012-01-16 & A24 TAC49 &  30  & 4 & 10 \\
ULASJ122343.35-013100.7 & CH$_4$s, CH$_4$l  & NICS         & 2012-04-29 & A25 TAC32 &  30  & 4 & 10 \\
\hline
\end{tabular}
\caption{Summary of observations
\label{tab:observations}
}
\end{table*}

\begin{table*}
\addtocounter{table}{-1}
\begin{tabular}{ c  c  c  c c c c c }
\hline
Object & Filter & Instrument & Date & TNG Program & T int (s) & N coadds & N dither \\
\hline
ULASJ122633.36+152106.9 & CH$_4$s, CH$_4$l  & NICS         & 2011-05-13 & A23 TAC28 &  30  & 2 & 30 \\
                                              & z                               & DOLORES & &  &  &  & -- \\
ULASJ122839.50+040758.5 & CH$_4$s, CH$_4$l  & NICS         & 2011-05-11 & A23 TAC28 &  30  & 1 & 30 \\
ULASJ124639.33+032314.3 & CH$_4$s, CH$_4$l  & NICS         & 2012-01-17 & A24 TAC49 &  30  & 4 & 10 \\
SDSS J124739.04+064604. & CH$_4$s, CH$_4$l  & NICS         & 2012-01-14 & A24 TAC49 &  30  & 4 & 10 \\
ULASJ125015.59+262846.8 & CH$_4$s, CH$_4$l  & NICS         & 2012-01-14 & A24 TAC49 &  20  & 3 & 10 \\
ULASJ125149.84+235653.8 & CH$_4$s, CH$_4$l  & NICS         & 2012-05-01 & A25 TAC32 &  30  & 4 & 10 \\
ULASJ125359.78+265855.5 & CH$_4$s, CH$_4$l  & NICS         & 2012-04-30 & A25 TAC32 &  30  & 4 & 10 \\
ULASJ125446.35+122215.7 & CH$_4$s, CH$_4$l  & NICS         & 2012-01-14 & A24 TAC49 &  30  & 4 & 10 \\
ULASJ125939.44+293322.4 & CH$_4$s, CH$_4$l  & NICS         & 2012-02-01 & A24 TAC49 &  30  & 4 & 10 \\
ULASJ130154.79+064747.9 & CH$_4$s, CH$_4$l  & NICS         & 2012-01-16 & A24 TAC49 &  30  & 4 & 10 \\
ULASJ130227.54+143428.0 & CH$_4$s, CH$_4$l  & NICS         & 2012-01-16 & A24 TAC49 &  30  & 4 & 10 \\
ULASJ130444.26+310112.3 & CH$_4$s, CH$_4$l  & NICS         & 2013-01-26 & A26 TAC68 &  26, 30  & 3, 4 & 30 \\
ULASJ130716.63+332523.8 & CH$_4$s, CH$_4$l  & NICS         & 2012-01-16 & A24 TAC49 &  30  & 4 & 10 \\
ULASJ131304.03+225919.7 & CH$_4$s, CH$_4$l  & NICS         & 2012-01-15 & A24 TAC49 &  30  & 4 & 10 \\
ULASJ131313.57+265433.8 & CH$_4$s, CH$_4$l  & NICS         & 2012-01-17 & A24 TAC49 &  30  & 4 & 10 \\
ULASJ131610.28+075553.0 & CH$_4$s, CH$_4$l  & NICS         & 2013-01-26 & A26 TAC68 &  26, 30  & 3, 4 & 30 \\
ULASJ131615.93+331306.8 & CH$_4$s, CH$_4$l  & NICS         & 2012-01-14 & A24 TAC49 &  30  & 4 & 10 \\
ULASJ131858.09-001632.3 & CH$_4$s, CH$_4$l  & NICS         & 2011-07-06 & A23 TAC28 &  30  & 2 & 30 \\
                                              & z                               & DOLORES & &  &  &  & -- \\
ULASJ132125.90+073549.4 & CH$_4$s, CH$_4$l  & NICS         & 2012-04-29 & A25 TAC32 &  30  & 4 & 10 \\
ULASJ133502.11+150653.5 & CH$_4$s, CH$_4$l  & NICS         & 2011-05-10 & A23 TAC28 &  30  & 1 & 30 \\
ULASJ133750.47+263648.7 & CH$_4$s, CH$_4$l  & NICS         & 2012-01-17 & A24 TAC49 &  20  & 3 & 10 \\
ULASJ134646.71+282009.2 & CH$_4$s, CH$_4$l  & NICS         & 2012-01-14 & A24 TAC49 &  30  & 3 & 10 \\
ULASJ134926.40+234045.9 & CH$_4$s, CH$_4$l  & NICS         & 2012-05-01 & A25 TAC32 &  30  & 4 & 10 \\
ULASJ135322.83+283408.2 & CH$_4$s, CH$_4$l  & NICS         & 2012-01-15 & A24 TAC49 &  30  & 4 & 10 \\
ULASJ135816.34+300539.1 & CH$_4$s, CH$_4$l  & NICS         & 2012-05-01 & A25 TAC32 &  30  & 4 & 10 \\
ULASJ141520.66+041647.1 & CH$_4$s, CH$_4$l  & NICS         & 2012-01-16 & A24 TAC49 &  30  & 4 & 10 \\
ULASJ141756.22+133045.8 & CH$_4$s, CH$_4$l  & NICS         & 2011-05-07 & A23 TAC28 &  30  & 1 & 30 \\
ULASJ142007.60+021818.3 & CH$_4$s, CH$_4$l  & NICS         & 2012-01-16 & A24 TAC49 &  30  & 4 & 10 \\
ULASJ142210.00+003023.7 & CH$_4$s, CH$_4$l  & NICS         & 2012-04-30 & A25 TAC32 &  30  & 4 & 10 \\
ULASJ142536.35+045132.3 & CH$_4$s, CH$_4$l  & NICS         & 2011-05-13 & A23 TAC28 &  30  & 2 & 30 \\
                                              & z                               & DOLORES & &  &  &  & -- \\
ULASJ144609.14+020300.3 & CH$_4$s, CH$_4$l  & NICS         & 2011-05-10 & A23 TAC28 &  30  & 2 & 30 \\
ULASJ144901.91+114711.4 & CH$_4$s, CH$_4$l  & NICS         & 2011-05-07 & A23 TAC28 &  30  & 1 & 30 \\
ULASJ151637.89+011050.1 & CH$_4$s, CH$_4$l  & NICS         & 2011-07-09 & A23 TAC28 &  30  & 1 & 30 \\
ULASJ153311.90-010612.9 & CH$_4$s, CH$_4$l  & NICS         & 2012-05-01 & A25 TAC32 &  30  & 4 & 10 \\
ULASJ153406.06+055643.9 & CH$_4$s, CH$_4$l  & NICS         & 2011-05-13 & A23 TAC28 &  30  & 2 & 30 \\
                                              & z                               & DOLORES & &  &  &  & -- \\
ULASJ153608.64+030556.5 & CH$_4$s, CH$_4$l  & NICS         & 2011-05-11 & A23 TAC28 &  30  & 2 & 30 \\
                                              & z                               & DOLORES & &  &  &  & -- \\
ULASJ154427.34+081926.6 & CH$_4$s, CH$_4$l  & NICS         & 2011-07-10 & A23 TAC28 &  30  & 1 & 30 \\
ULASJ154914.45+262145.6 & CH$_4$s, CH$_4$l  & NICS         & 2011-07-10 & A23 TAC28 &  30  & 1 & 30 \\
ULASJ155250.22+013606.6 & CH$_4$s, CH$_4$l  & NICS         & 2011-07-08 & A23 TAC28 &  30  & 2 & 50 \\
                                              & z                               & DOLORES & &  &  &  & -- \\
ULASJ161436.96+244230.1 & CH$_4$s, CH$_4$l  & NICS         & 2011-05-10 & A23 TAC28 &  30  & 1 & 30 \\
                                              & z                               & DOLORES & &  &  &  & -- \\
ULASJ161710.39+235031.4 & CH$_4$s, CH$_4$l  & NICS         & 2011-05-07 & A23 TAC28 &  30  & 1 & 30 \\
ULASJ161934.78+235829.3 & CH$_4$s, CH$_4$l  & NICS         & 2011-07-11 & A23 TAC28 &  30  & 2 & 30 \\
ULASJ161938.12+300756.4 & CH$_4$s, CH$_4$l  & NICS         & 2011-05-08 & A23 TAC28 &  30  & 2 & 30 \\
ULASJ203920.56+002638.3 & CH$_4$s, CH$_4$l  & NICS         & 2012-08-02 & A26 TAC68 &  26, 30  & 3, 4 & 30 \\
ULASJ211317.05+001840.7 & CH$_4$s, CH$_4$l  & NICS         & 2011-10-27 & A24 TAC49 &  30  & 1 & 30 \\
ULASJ211616.26-010124.3  & CH$_4$s, CH$_4$l  & NICS         & 2011-10-26 & A24 TAC49 &  45  & 1 & 30 \\
                                              & z                               & DOLORES & &  &  &  & -- \\
ULASJ213352.64-010343.4  & CH$_4$s, CH$_4$l  & NICS         & 2012-04-29 & A25 TAC32 &  30  & 4 & 10 \\
                                              & z                               & DOLORES & &  &  &  & -- \\
ULASJ214112.84-010954.6 & CH$_4$s, CH$_4$l  & NICS         & 2011-07-14 & A23 TAC28 &  30  & 2 & 50 \\
                                              & z                               & DOLORES & &  &  &  & -- \\
ULASJ215343.25-001626.0 & CH$_4$s, CH$_4$l  & NICS         & 2011-10-27 & A24 TAC49 &  60  & 1 & 30 \\
                                              & z                               & DOLORES & &  &  &  & -- \\
ULASJ221606.39+032159.2 & CH$_4$s, CH$_4$l  & NICS         & 2011-10-26 & A24 TAC49 &  60  & 1 & 30 \\
WISE J222623.05+044003. & CH$_4$s, CH$_4$l  & NICS         & 2011-11-18 & A24 TAC49 &  30  & 1 & 30 \\
ULASJ223748.61+052039.6 & CH$_4$s, CH$_4$l  & NICS         & 2011-10-27 & A24 TAC49 &  30  & 1 & 30 \\
ULASJ223917.13+073416.0 & CH$_4$s, CH$_4$l  & NICS         & 2011-07-09 & A23 TAC28 &  30  & 2 & 30 \\
                                              & z                               & DOLORES & &  &  &  & -- \\
\hline
\end{tabular}
\caption{Summary of observations (continued)
\label{tab:observations}
}
\end{table*}

\begin{table*}
\addtocounter{table}{-1}
\begin{tabular}{ c  c  c  c c c c c }
\hline
Object & Filter & Instrument & Date & TNG Program & T int (s) & N coadds & N dither \\
\hline

ULASJ225023.52-001605.9 & CH$_4$s, CH$_4$l  & NICS         & 2011-07-14 & A23 TAC28 &  30  & 2 & 30 \\
                                              & z                               & DOLORES & &  &  &  & -- \\
ULASJ225540.22+061412.9 & CH$_4$s, CH$_4$l  & NICS         & 2011-07-09 & A23 TAC28 &  30  & 2 & 30 \\
                                              & z                               & DOLORES & &  &  &  & -- \\
ULASJ230049.08+070338.0 & CH$_4$s, CH$_4$l  & NICS         & 2011-07-09 & A23 TAC28 &  30  & 1 & 30 \\
                                              & z                               & DOLORES & &  &  &  & -- \\
ULASJ232600.40+020139.2 & CH$_4$s, CH$_4$l  & NICS         & 2011-10-27 & A24 TAC49 &  45  & 1 & 30 \\
ULASJ232624.07+050931.6 & CH$_4$s, CH$_4$l  & NICS         & 2012-01-16 & A24 TAC49 &  30  & 4 & 10 \\
                                              & z                               & DOLORES & &  &  &  & -- \\
ULASJ235204.62+124444.9 & CH$_4$s, CH$_4$l  & NICS         & 2011-10-28 & A24 TAC49 &  30  & 1 & 30 \\

\hline
\end{tabular}
\caption{Summary of observations (continued)
\label{tab:observations}
}
\end{table*}

\section*{Acknowledgements}
The author's would like to thank the staff of the Telescopio Nazionale Galileo (TNG) for their excellent support during of the very large program on which this work is based. 
BB acknowledges financial support from the European
Commission in the form of a Marie Curie International Outgoing Fellowship (PIOF-GA-2013-629435).
This research has benefitted from the SpeX Prism Spectral Libraries, maintained by Adam Burgasser at 
\url{http://pono.ucsd.edu/~adam/browndwarfs/spexprism}
The authors also acknowledge the Marie Curie 7th European
Community Framework Programme grant n.236735 Parallaxes of
Southern Extremely Cool objects (PARSEC) International Incoming
Fellowship and grant n.247593  Interpretation and Parameterization
of Extremely Red COOL dwarfs (IPERCOOL) International Research
Staff Exchange Scheme. Based on observations made with the Italian
Telescopio Nazionale Galileo operated on the island of La Palma by the
Fundaci—n Galileo Galilei of the Istituto Nazionale di
Astrofisica at the Spanish Observatorio del Roque de los Muchachos of
the Instituto de Astrofisica de Canarias.
The United Kingdom Infrared Telescope (UKIRT) was operated by the Joint Astronomy
Centre on behalf of the Science and Technology Facilities Council of the
U.K.
This work is based in part on data obtained as part of the UKIRT Infrared Deep Sky Survey.

\bibliographystyle{mn2e}
\bibliography{CH4}

\end{document}